\documentclass[11pt,a4paper]{article}

\usepackage{graphicx,epsf,epsfig,amssymb,amsmath,amsfonts,mathrsfs}
\usepackage{longtable}
\usepackage{bm}
\usepackage{color}
\usepackage{enumitem}
\usepackage{float}
\usepackage{caption}
\usepackage{graphicx}
\usepackage{booktabs}
\usepackage{array}
\usepackage{tabularx}
\usepackage{array}
\newcolumntype{Y}{>{\centering\arraybackslash}X}
\usepackage{enumitem}
\usepackage{caption}
\usepackage{subcaption}
\usepackage[section]{placeins}

\usepackage{subcaption}
\usepackage{upgreek}
\usepackage{cite}
\usepackage{mathtools}
\usepackage{titlesec}
\usepackage{lipsum}
\usepackage{xcolor}
\usepackage[section]{placeins}
\usepackage[colorlinks=true,linktocpage=true,linkcolor=blue,citecolor=blue]{hyperref}
\usepackage{tensor}
\usepackage{multirow}
\usepackage[utf8]{inputenc}
\usepackage{authblk}
\usepackage{array}

\usepackage{tikz}
\usepackage{booktabs}
\usepackage[normalem]{ulem}
\usepackage{soul}
\usepackage[toc,page]{appendix}

\newcommand{\be}{\begin{equation}}
\newcommand{\ee}{\end{equation}}
\newcommand{\bea}{\begin{eqnarray}}
\newcommand{\eea}{\end{eqnarray}}

\newcommand{\dmoff}[1]{}

\newcommand\blfootnote[1]{%
  \begingroup
  \renewcommand\thefootnote{}\footnote{#1}%
  \addtocounter{footnote}{-1}%
  \endgroup
}

\usepackage{comment}

\newcommand{\customrule}{%
\vspace{0.5cm}
\noindent\textcolor{gray}{\rule{\textwidth}{0.5pt}}
\vspace{0.5cm}
}

\usepackage[
a4paper,
left=1.5cm,
right=1.5cm,
top=3cm,
bottom=4cm,
]{geometry}

\numberwithin{equation}{section} 

\titleclass{\subsubsubsection}{straight}[\subsubsection]
\newcounter{subsubsubsection}[subsubsection]
\renewcommand\thesubsubsubsection{\thesubsubsection.\arabic{subsubsubsection}}

\titleformat{\subsubsubsection}[block]
  {\normalfont\normalsize\bfseries}
  {\thesubsubsubsection}{1em}{}

\titlespacing*{\subsubsubsection}
  {0pt}{3.25ex plus 1ex minus .2ex}{1em}

\makeatletter
\def\toclevel@subsubsubsection{4}
\def\l@subsubsubsection{\@dottedtocline{4}{7em}{4em}}
\makeatother

\author[1]{Etevaldo dos Santos Costa Filho}
\author[2,3]{Romain Gervalle}

\affil[1]{\normalsize Programa de Pós-Graduação em Física, Universidade Federal do Espírito Santo, Vitória, ES,  29075-910, Brazil}
\affil[2]{\normalsize Departamento de Matemática da Universidade de Aveiro and Center for Research and Development in Mathematics and Applications -- CIDMA

Campus de Santiago, 3810-183 Aveiro, Portugal}
\affil[3]{\normalsize Département de Physique de l'Université de Tours, Parc de Grandmont, 37200 Tours, France}

\begin{document}
\title{\bf Bosonic stars with dark electroweak fields}

\maketitle

\begin{abstract}
We construct  bosonic stars in the 
Einstein-Weinberg-Salam theory, namely the bosonic sector of the electroweak Standard Model minimally coupled to Einstein's theory of gravity. These configurations are everywhere regular, spherically symmetric and asymptotically flat, consisting of a static condensate of the massive $W$ and $Z$ bosons, whose masses arise from the Higgs mechanism. Although the corresponding fields exist in nature, gravitational effects are negligible at the physical electroweak scale  and the configurations would be microscopic. 
Keeping Newton's constant and the electroweak mass ratios at their physical values, we therefore interpret the model as the bosonic sector of a \textit{dark} electroweak theory with a radically different energy scale. Choosing the ultralight vector boson mass $m_{\mathrm{W}}c^{2}=8.7\times10^{-13}\,$eV, as suggested by the Proca star interpretation of GW190521, 
our bosonic stars can attain masses in the intermediate-mass black hole range.
\end{abstract}
\vfill
\blfootnote{ {\tt etevaldo.s.costa@ufes.br}}
\blfootnote{ {\tt romaingervalle@ua.pt}}

	\newpage
	\customrule
	\tableofcontents

\customrule

\section{Introduction}

The problem of dark matter is now a century old \cite{Kapteyn1922,Oort1932,Zwicky1933}, and determining its fundamental nature remains a central open question at the interface of astrophysics and high-energy physics. Resolving this question may open a window into physics beyond the Standard Model. Indeed, growing evidence points toward dark matter being a new form of matter, rather than a
manifestation of modified gravity \cite{Markevitch2004,Klypin:2007gc,Clowe:2006eq}. However, collider experiments have not yet revealed the nature of these putative dark matter particles, 
motivating a complementary strategy: rather than searching only for their direct
production or scattering in laboratory experiments, one can also use compact objects and gravitational-waves to
probe dark sectors. Recent developments lend concrete support to this perspective.

In synergy with high‑energy physics, we may view the current gravitational‑wave detectors LIGO-Virgo-KAGRA as indirect particle detectors. A concrete example of this possible interplay is the gravitational‑wave event GW190521 \cite{GW190521}, which is conventionally interpreted as the merger of two Kerr black holes followed by the formation of a more massive remnant. However, the analysis in Ref.~\cite{CalderonBustillo:2020fyi} showed that the same
signal admits an alternative interpretation: GW190521 may have been produced by the head-on collision of two Proca stars \cite{Brito:2015pxa} -- horizonless, self-gravitating 
condensates of a massive complex vector field~-- that subsequently collapse into a final black hole. Under this interpretation, the hypothetical vector boson would be ultralight, with a mass $\simeq8.7\times10^{-13}\,\mathrm{eV}/c^{2}$. In a subsequent work \cite{CalderonBustillo:2022cja}, other peculiar gravitational‑wave signals detected by LIGO and Virgo were compared with Proca‑star merger waveforms, revealing again a degeneracy with the usual binary black hole interpretation. 
Remarkably, this study not only reported the possibility that Proca stars explain certain astrophysical events, but also found a boson mass consistent across different events, further supporting the potential astrophysical role of these exotic compact objects. 

Proca stars belong to the larger family of bosonic stars, namely, horizonless, self-gravitating configurations composed of scalar or vector fields, which have been studied for more than fifty years\cite{Kaup:1968zz,Ruffini:1969qy,Jetzer:1991jr,Schunck:2003kk,Liebling:2012fv,Shnir:2022lba,Schunck:1998cdq,Mielke:2000mh,Yuan:2004sv,Guzman:2005bs,Berti:2006qt,Guzman:2009zz,Vincent:2015xta,Sennett:2017etc,Grould:2017rzz,Olivares:2018abq,Herdeiro:2021lwl,Rosa:2022tfv,Rosa:2023qcv,Herdeiro:2023wqf,Adam:2024zqr,Herdeiro:2026agu,Diez-Tejedor:2026fnc,Colpi:1986ye}. Even though the existence of such stars is speculative, in some scenarios they can act as \textit{black hole mimickers}, offering a useful framework for developing concepts and tools that help us deal with degeneracies between observations and different theoretical models 
\cite{Herdeiro:2021lwl,Olivares:2018abq,Sengo:2022jif,Rosa:2022tfv,Rosa:2023qcv,CalderonBustillo:2020fyi,Sanchis-Gual:2018oui,CalderonBustillo:2022cja,Palenzuela:2006wp,herdeiro2023procahiggs,Brito:2024biy,Herdeiro:2024pmv,Atteneder:2023pge,Tsukada:2020lgt,Lee:1995af,Suarez:2013iw,Eby:2015hsq,Chen:2020cef,Jones:2023fzz,Ryan:1996nk,ryan1995gravitational,Siemonsen:2023age}. In parts of their parameter space, these stars are dynamically robust and can be evolved in binary systems using numerical techniques \cite{Liebling:2012fv,Sanchis-Gual:2018oui}, yielding exotic but consistent models, whose gravitational-wave signals can be directly confronted with observations~\cite{CalderonBustillo:2020fyi,CalderonBustillo:2022cja,Sanchis-Gual:2022mkk}.  Bosonic stars exist both in General Relativity (GR) and in modified theories of gravity, but they typically require matter content beyond the Standard Model. In the simplest models where the bosonic field is minimally coupled to gravity and lacks self-interactions, 
achieving astrophysical masses for the stars requires the field mass  to lie in the ultralight range $10^{-20}\!-\!10^{-10}\,\mathrm{eV}/c^2$. From a particle physics perspective, such fields can be interpreted as a form of diffuse dark matter, a scenario that has recently attracted considerable attention \cite{Hui2017,Antypas:2022asj,Ferreira:2020fam}.

However, even the free Proca model should be regarded as an effective field theory in the context of high-energy physics.
Indeed, the fixed mass term explicitly breaks gauge invariance, compromising the theoretical consistency of the model at high energies.
In a more fundamental description, 
one expects this mass term to be generated via a Higgs mechanism, as it is for the $W$ and $Z$ bosons, the massive vector bosons in the electroweak sector of the Standard Model. Moreover, the pure Proca model exhibits pathologies upon inclusion of self-interactions, including hyperbolicity violations and instabilities \cite{Minamitsuji:2018kof,Herdeiro:2020jzx,Coates:2022nif,Mou:2022hqb,PhysRevLett.129.151103,Barausse:2022rvg,PhysRevLett.129.151102}, which do not arise in the scalar case \cite{baer2008wave} and further signal its effective nature. Despite these issues, self-interacting Proca fields continue to be studied in the context of non-topological solitons and self-gravitating solutions, while generalized Proca models and theories involving vector-tensor couplings have been explored in cosmology and black hole physics \cite{Esposito-Farese:2009wbc,Annulli:2019fzq,Barton:2021wfj,Loginov:2015rya,Brihaye:2017inn,Charmousis:2025jpx,Fernandes:2026rjs}. It is worth noting, however, that the free massive Proca model remains hyperbolic \cite{Baer2015}.

On the other hand, a consistent effective‑field‑theory approach to such models requires embedding them in extensions of the Standard Model. In the specific case of ultralight bosons, one embedding has been proposed via the so‑called axiverse \cite{Arvanitaki2010,Cicoli:2012sz,Demirtas:2018akl}. Independently, simple extensions of the Standard Model yielding ultralight vectors or scalars with potential astrophysical relevance have also been considered \cite{Freitas:2021cfi}. In particular, a non-Abelian dark sector governed by the electroweak gauge group ${\rm SU}(2)_{\rm L}\times{\rm U}(1)_{\rm Y}$ provides a natural framework in which massive vector fields arise through a Higgs mechanism rather than through explicit Proca mass terms. Motivated by this, we consider in this work the bosonic sector of a \textit{dark} electroweak theory minimally coupled to GR. We assume Standard-Model-like boson mass ratios, but an overall energy scale that differs radically from its observed electroweak value. Within this setting, we construct everywhere regular, asymptotically flat and spherically symmetric solutions describing new bosonic stars. We note that closely related self-gravitating solitons in ${\rm SU}(2)$ Yang-Mills theory were studied in Refs.~\cite{Jain:2022kwq,Bartnik:1988am,Brihaye:2004nd}.

The qualifier \textit{dark} is not optional if one intends to construct objects of astrophysical relevance. At the observed electroweak scale, the dimensionless coupling measuring the gravitational backreaction of the electroweak fields is of
order $\sim10^{-33}$, so that the corresponding self-gravitating configurations would be microscopic, with negligible
gravitational effects. Keeping Newton's constant at its physical value and raising the Higgs vacuum expectation value of the \textit{dark} electroweak sector near the Planck scale brings this coupling to order unity, while choosing ultralight boson masses allows the solutions to describe astrophysical objects.
In particular, for a dark $W$ boson with
$m_{\rm W}=8.7\times10^{-13}\,\text{eV}/c^2$ -- the value suggested by the Proca star interpretation of
GW190521 \cite{CalderonBustillo:2020fyi,CalderonBustillo:2022cja} -- the configurations constructed here
reach masses up to $\sim100-1000\,M_{\odot}$.

We note that other classical configurations in the gravity-coupled electroweak theory are known. Gravitating generalizations of the Klinkhamer-Manton sphaleron~\cite{Klinkhamer1984} -- both regular and black hole solutions~--~were constructed in Ref.~\cite{Greene:1992fw}, albeit in the special limit of vanishing weak mixing angle $\theta_{\rm W}$ where the ${\rm U}(1)_{\rm Y}$ hypercharge field decouples. For the Standard-Model value of $\theta_{\rm W}$, sphalerons exist but break spherical symmetry~\cite{Kleihaus1991,Kunz1992}. More recently, solutions describing ``black holes inside magnetic monopoles'' have been constructed for the physical $\theta_{\rm W}$, first in the special case of spherical symmetry~\cite{Bai2021}, and later in the more general axially-symmetric case~\cite{Gervalle2024,Gervalle2025}. These are gravitating counterparts of the electroweak (multi)monopoles~\cite{Cho1996,Gervalle2022a,Gervalle2023}. In all of these solutions, however, the nontrivial field structure originates from the magnetic sector of the theory. By contrast, the solutions constructed here belong to the purely electric sector, which, to the best of our knowledge, has not been previously explored.

The rest of the paper is organized as follows. Section~\ref{sec:model} introduces the Einstein-Weinberg-Salam model and its dimensionless formulation. In Sec.~\ref{sec:ansatz}, we present the spherically symmetric field ansatz, define the relevant physical quantities, and describe our numerical methods. Section~\ref{Sec:solutions} presents the solutions and analyzes their main physical properties, while Sec.~\ref{sec:numerical_physical} relates the dimensionless numerical results to physical quantities, and clarifies their dark-electroweak interpretation. We conclude with further remarks in Sec.~\ref{sec:remarks}. Additionally, a Smarr-type mass formula is derived in Appendix~\ref{sec:massformula}.

\section{The Einstein-Weinberg-Salam model}\label{sec:model}

We consider the bosonic sector of the electroweak theory, with gauge group ${\rm SU}(2)_{\rm L}\times{\rm U}(1)_{\rm Y}$ (the Weinberg-Salam model \cite{Weinberg:1967tq,Salam:1968rm}), minimally coupled to Einstein's theory of gravity. Throughout the paper, we use the metric signature $(-,+,+,+)$. Boldface symbols denote dimensionful quantities, whereas ordinary symbols denote their dimensionless counterparts. With these conventions, the dimensionful action is,
\begin{equation}
\bm{S}
=
\frac{1}{\bm{c}}
\int
\left(
\frac{\bm{c}^{4}}{16\pi \bm{G}}\,\bm{R}
+
\bm{L}_{\mathrm{WS}}
\right)
\sqrt{-{\rm g}}\, d^{4}{\bm x},
\label{eq:action-dimensional}
\end{equation}
where $\bm c$ is the speed of light, $\bm G$ is Newton's constant, $\bm R$ is the Ricci scalar, and ${\bm L}_\text{WS}$ is the electroweak Lagrangian,
\begin{equation}
\bm{L}_{\mathrm{WS}}
=
-\frac{1}{4}\bm{W}^{a}_{\mu\nu}\bm{W}^{a\mu\nu}
-\frac{1}{4}\bm{B}_{\mu\nu}\bm{B}^{\mu\nu}
-({\bm D}_{\mu}\bm{\Phi})^{\dagger}{\bm D}^{\mu}\bm{\Phi}
-{\bm\lambda}\left(\bm{\Phi}^{\dagger}\bm{\Phi}-\bm{\Phi}_{0}^{2}\right)^{2}.
\label{eq:lagrangian-ws-dimensional}
\end{equation}

The latter contains the ${\rm SU}(2)$ and ${\rm U}(1)$ field strengths given by,
\begin{equation}
\bm{W}^{a}_{\mu\nu}
=
\bm{\partial}_{\mu}\bm{W}^{a}_{\nu}
-
\bm{\partial}_{\nu}\bm{W}^{a}_{\mu}
+
\bm{g}\,\epsilon_{abc}\,\bm{W}^{b}_{\mu}\bm{W}^{c}_{\nu},
\qquad
\bm{B}_{\mu\nu}
=
\bm{\partial}_{\mu}\bm{B}_{\nu}
-
\bm{\partial}_{\nu}\bm{B}_{\mu},
\label{eq:field-strengths-dimensional}
\end{equation}
with $\bm{\partial}_\mu=\partial/\partial{\bm x}^\mu$. The Higgs field is in the fundamental representation of the group ${\rm SU}(2)$, hence it is a complex doublet ${\bm\Phi}=(\bm \phi_1, \bm \phi_2)^T$ with the gauge covariant derivative,
\begin{equation}
{\bm D}_{\mu}\bm{\Phi}
=
\left(
{\bm\partial}_{\mu}
-\frac{i\bm{g}'}{2}\bm{B}_{\mu}
-\frac{i\bm{g}}{2}\tau_{a}\bm{W}^{a}_{\mu}
\right)\bm{\Phi},
\label{eq:higgs-covariant-dimensional}
\end{equation}
where $\tau_a$ are the Pauli matrices. The Lagrangian
\eqref{eq:lagrangian-ws-dimensional} contains four parameters: the gauge
couplings $\bm g$, $\bm g'$, the Higgs self-coupling $\bm \lambda$, and the
Higgs vacuum expectation value $\bm \Phi_0$. After symmetry breaking, ${\rm SU}(2)_{\rm L}\times{\rm U}(1)_{\rm Y} \to
{\rm U}(1)_{\rm em}$, three of the four gauge bosons acquire masses via the
Higgs mechanism, while the photon remains massless. 

Let us now introduce dimensionless quantities,
\begin{equation}
W^{a}_{\mu}
=
\frac{g}{\bm{\Phi}_{0}}\,\bm{W}^{a}_{\mu},
\quad
B_{\mu}
=
\frac{g'}{\bm{\Phi}_{0}}\,\bm{B}_{\mu},
\quad
\Phi
=
\frac{\bm{\Phi}}{\bm{\Phi}_{0}},
\quad
\beta
=
\frac{8{\bm\lambda}}{{\bm g}_{0}^{2}},
\quad
g
=
\frac{\bm{g}}{{\bm g}_{0}},
\quad
g'
=
\frac{\bm{g}'}{{\bm g}_{0}},
\quad
x^{\mu}
=
\bm{x}^{\mu}{\bm g}_0{\bm\Phi}_{0},
\label{eq:dimensionless-variables}
\end{equation}
where ${\bm g}_0=\sqrt{{\bm g}^2+{\bm g}'^2}$. We also introduce the dimensionless gravitational coupling,
\begin{equation}
    \kappa = \frac{8\pi \bm{G}\,\bm{\Phi}_0^2}{\bm{c}^4},
\label{eq:kappa}
\end{equation}
and finally define the dimensionless action $S\equiv{\bm S}/{\bm\hbar}$, whose explicit expression is, 
\begin{align}
S &= \frac{e^{2}}{4\pi\alpha}\int
\left(
\frac{1}{2\kappa}R + \mathcal{L}_{\mathrm{WS}}
\right)\sqrt{-{\rm g}}\,d^{4}x,
\label{eq:action-dimensionless}
\\
\mathcal{L}_{\mathrm{WS}}
&=
-\frac{1}{4g^{2}}\,W^{a}_{\mu\nu}W^{a\mu\nu}
-\frac{1}{4g'^{2}}\,B_{\mu\nu}B^{\mu\nu}
-(D_{\mu}\Phi)^{\dagger}D^{\mu}\Phi
-\frac{\beta}{8}\left(\Phi^{\dagger}\Phi-1\right)^{2}.
\nonumber
\end{align}
Here we have introduced the (dimensionless) elementary charge $e\equiv gg'$ and the fine structure constant $\alpha\equiv\, e^{2}\bm\hbar\bm c\bm g_0^2/(4\pi )$. The field strengths and gauge covariant derivative now read,
\begin{equation}
    W^a_{\mu\nu}=\partial_\mu W^a_\nu-\partial_\nu W^a_\mu+\epsilon_{abc}W^b_\mu W^c_\nu,\quad\quad B_{\mu\nu}=\partial_\mu B_\nu-\partial_\nu B_\mu,\quad\quad D_\mu\Phi=\left(\partial_\mu-\frac{i}{2}B_\mu-\frac{i}{2}\tau_a W^a_\mu\right)\Phi.
\end{equation}

The action \eqref{eq:action-dimensionless} describes a theory of the complex-valued Higgs field $\Phi=(\phi_1,\phi_2)^T$, the U(1) hypercharge field $B=B_\mu dx^\mu$, and the ${\rm SU}(2)$ field $W=T_a W^a_\mu dx^\mu$ (with $T_a=\tau_a/2$, the ${\rm SU}(2)$ generators), all interacting with the gravitational metric field $\rm g_{\mu\nu}$. The overall factor, $e^{2}/(4\pi\alpha)=1/(\bm\hbar\bm c\bm g_0^2)$, does not appear in the field equations and merely normalizes the action. However, its magnitude controls the accuracy of the classical description for the fields. For Standard Model couplings it is of order unity, such that $\bm S \sim {\bm \hbar}$, whereas for the coupling values considered in this paper (see Sec.~\ref{sec:numerical_physical}) it becomes enormous and $\bm S\gg {\bm \hbar}$. Hence, the configurations constructed in this work contain a macroscopic number of quanta and are deep in the classical regime \cite{Herdeiro:2022gzp}.

We now adopt the unitary gauge, where the Higgs field reduces to a single scalar component, $\Phi=(0,\phi)^T$, and introduce a complex vector field $w_\mu$ along with its field strength,
\begin{equation}
    w_\mu=\frac{1}{g}(W^1_\mu + i W^2_\mu),\quad\quad w_{\mu\nu}={\cal D}_\mu w_\nu -{\cal D}_\nu w_\mu,
\label{eq:w-proca-field}
\end{equation}
where ${\cal D}_\mu=\nabla_\mu +iW^3_\mu$. Next, we define the electromagnetic and $Z$ fields following 't Hooft (see Ref.~\cite{tHooft:1974kcl,Nambu:1977ag,Gervalle2025} for details and the comparison with the alternative definition due to Nambu),
\begin{equation}
    A_\mu=\frac{g}{g'}B_\mu+\frac{g'}{g}W^3_\mu,\qquad Z_\mu=B_\mu-W^3_\mu,
\end{equation}
with the corresponding field strengths,
\begin{equation}
    F_{\mu\nu}=\partial_\mu A_\nu-\partial_\nu A_\mu,\qquad Z_{\mu\nu}=\partial_\mu Z_\nu-\partial_\nu Z_\mu.
\end{equation}
The calligraphic covariant derivative now takes the form,
\begin{equation}
    {\cal D}_\mu=\nabla_\mu+i(e A_\mu-g^2 Z_\mu).
\end{equation}

Expressed in terms of these new fields, the electroweak Lagrangian in Eq.~\eqref{eq:action-dimensionless} becomes,
\begin{align}
\mathcal{L}_{\mathrm{WS}}
={}& -\frac{1}{4}F_{\mu\nu}F^{\mu\nu}
- \frac{1}{4}Z_{\mu\nu}Z^{\mu\nu}
- \frac{g g'}{2}F_{\mu\nu}\psi^{\mu\nu}
+ \frac{g^2}{2}Z_{\mu\nu}\psi^{\mu\nu}
-\frac{1}{4} w_{\mu\nu}\bar w^{\mu\nu} \notag\\
&-\frac{g^2}{4}\psi_{\mu\nu}\psi^{\mu\nu} 
- \partial_\mu\phi\,\partial^\mu\phi
- \frac{1}{4}\left(g^2 w_\mu \bar w^\mu + Z_\mu Z^\mu\right)\phi^2
- \frac{\beta}{8}\left(\phi^2 - 1\right)^2 \label{eq:Lagr},
\end{align}
where we have introduced the real antisymmetric  tensor,
\begin{equation}
    \psi_{\mu\nu}=\frac{i}{2}(w_\mu \bar{w}_\nu-w_\nu \bar{w}_\mu).
\end{equation}

Although lengthier, this form of the Lagrangian makes the particle spectrum completely transparent. The massless electromagnetic field 
$A_\mu$ couples to the charged vector field $w_\mu$ describing the $W$ bosons, both through the covariant derivative ${\cal D}_\mu$ and through the tensor $\psi_{\mu\nu}$. The field $Z_{\mu}$ acquires a mass through its coupling to the Higgs, and is sourced by $w_\mu$ in the same fashion as 
$A_\mu$. The complex field $w_\mu$ exhibits quartic self-interactions through the term $\psi_{\mu\nu}\psi^{\mu\nu}$, and likewise acquires a mass via its coupling to the Higgs. Finally, the massive Higgs field $\phi$ is governed by a Mexican hat potential, giving rise to a nonzero vacuum expectation value (vev), which in our conventions~\eqref{eq:dimensionless-variables} is normalized to unity.

Apart from general covariance, the Lagrangian \eqref{eq:Lagr} possesses the residual ${\rm U}(1)_{\rm em}$ gauge symmetry of the spontaneously broken electroweak sector,
\begin{align}\label{eq:residualU1}
    {\rm U}(1)_{\rm em}\; :\quad w_\mu\to e^{-i\lambda(x)}w_\mu,\quad \bar{w}_\mu\to e^{i\lambda(x)}\bar{w}_\mu,\quad A_\mu\to A_\mu+\frac{1}{e}\partial_\mu\lambda(x),\quad Z_\mu\to Z_\mu,\quad \phi\to\phi.
\end{align}

In particular, the passage from the fields $(B_\mu,W^3_\mu)$ to $(A_\mu,Z_\mu)$, together with the unitary-gauge elimination of the Goldstone modes, fixes the original ${\rm SU}(2)_{\rm L}\times{\rm U}(1)_{\rm Y}$ gauge redundancy so that it is no longer manifest in the variables $(A_\mu,Z_\mu,w_\mu,\bar{w}_\mu,\phi)$. The field $A_\mu$ is the connection of the unbroken electromagnetic ${\rm U}(1)_{\rm em}$ subgroup, whereas $Z_\mu$, being a neutral massive vector, does not define an independent gauge symmetry. Correspondingly, \(w_\mu\) and \(\bar w_\mu\) transform covariantly, with opposite local phases. The derivative
${\cal D}_\mu=\nabla_\mu+i(eA_\mu-g^2 Z_\mu)$ is constructed precisely so
that ${\cal D}_\mu w_\nu$, and hence the field strength $w_{\mu\nu}$,
transforms covariantly, i.e., with the same phase as $w_\mu$. The bilinears
$\bar w_{\mu\nu}w^{\mu\nu}$, $w_\mu\bar w^\mu$ and $\psi_{\mu\nu}$, as well
as $F_{\mu\nu}$ and $Z_{\mu\nu}$, are then invariant, so that every term in
\eqref{eq:Lagr} is invariant under ${\rm U}(1)_{\rm em}$. The Lagrangian \eqref{eq:Lagr}
should therefore be regarded as the unitary-gauge form of the original
${\rm SU}(2)_{\rm L}\times{\rm U}(1)_{\rm Y}$ theory in Eq.~\eqref{eq:action-dimensionless}, with
${\rm U}(1)_{\rm em}$ as the only manifest internal gauge symmetry.

Varying the action \eqref{eq:action-dimensionless} with respect to $A_\mu$, $Z_\mu$, $w_\mu$ and $\phi$ gives the electroweak equations,
\begin{align}
    \nabla_\mu F^{\nu\mu}&=gg' J^\nu,
    \label{eq:A} \\
    \nabla_\mu Z^{\nu\mu}+\frac{1}{2}\phi^2 Z^\nu&=-g^2 J^\nu,
    \label{eq:Z} \\
    {\cal D}_\mu w^{\mu\nu}+i\left(g^2\psi^{\nu\sigma}+gg' F^{\nu\sigma}-g^2 Z^{\nu\sigma}\right)w_\sigma&=\frac{g^2}{2}\phi^2 w^\nu, 
    \label{eq:W} \\
    \nabla_\mu\nabla^\mu\phi-\frac{1}{4}\left(g^2 w_\mu\bar{w}^\mu+Z_\mu Z^\mu\right)\phi&=\frac{\beta}{4}(\phi^2-1)\phi,
    \label{eq:phi}
\end{align}
where we have introduced the electric current,
\begin{equation}\label{eq:current}
    J^\nu=\nabla_\mu\psi^{\mu\nu}+\frac{i}{2}\left(w_\mu\bar{w}^{\mu\nu}-\bar{w}_\mu w^{\mu\nu}\right),
\end{equation}
which is conserved by virtue of the Maxwell-like equation \eqref{eq:A}.
Notice that the $Z$ field is sourced by the neutral current proportional to $-J^\nu$, see Eq.~\eqref{eq:Z}, so that the electromagnetic and $Z$ fields are sourced by oppositely directed currents. Then, varying the action with respect to $\rm g_{\mu\nu}$ yields the Einstein equations,
\begin{equation}
    R_{\mu\nu}-\frac{1}{2}{\rm g}_{\mu\nu}R=\kappa\,T_{\mu\nu},
    \label{eq:Ein}
\end{equation}
where $R_{\mu\nu}$ is the Ricci tensor, and $T_{\mu\nu}$ is the electroweak stress-energy tensor whose explicit expression is,
\begin{align}
    T_{\mu\nu}=&\; F_{\mu\sigma}\tensor{F}{_\nu^\sigma} + Z_{\mu\sigma}\tensor{Z}{_\nu^\sigma} + gg'\left(F_{\mu\sigma}\tensor{\psi}{_\nu^\sigma}+F_{\nu\sigma}\tensor{\psi}{_\mu^\sigma}\right)-g^2\left(Z_{\mu\sigma}\tensor{\psi}{_\nu^\sigma}+Z_{\nu\sigma}\tensor{\psi}{_\mu^\sigma}\right)+g^2\psi_{\mu\sigma}\tensor{\psi}{_\nu^\sigma}\notag\\
    &+\frac{1}{2}\left(\bar{w}_{\mu\sigma}\tensor{w}{_\nu^\sigma}+\bar{w}_{\nu\sigma}\tensor{w}{_\mu^\sigma}\right)+2\partial_\mu\phi\,\partial_\nu\phi+\frac{1}{4}g^2\phi^2\left(\bar{w}_\mu w_\nu+\bar{w}_\nu w_\mu\right)+\frac{1}{2}\phi^2 Z_\mu Z_\nu+{\rm g}_{\mu\nu}\mathcal{L}_{\mathrm{WS}}.
\end{align}

One has the usual on-shell energy conservation condition,
\begin{equation}
    \nabla_\mu T^{\mu\nu}=0.
\end{equation}

Additionally, taking the gauge-covariant divergence of Eq.~\eqref{eq:W} and the covariant divergence of Eq.~\eqref{eq:Z} yields two Lorenz-type constraints,
\begin{align}
{\cal D}^{\mu}\!\left(\phi^{2}w_{\mu}\right)&=-\,i\,\phi^{2}Z^{\mu}w_{\mu},
\label{eq:omega-Lorenz}\\[2pt]
\nabla_{\mu}\!\left(\phi^{2}Z^{\mu}\right)&=0.
\label{eq:Zlorenz}
\end{align}

These are constraints rather than independent dynamical equations. They follow from  the antisymmetry of the field strengths $w_{\mu\nu}$ and $Z_{\mu\nu}$ together with current conservation.

The vacuum is the state with $T_{\mu\nu}=0$ and, up to a gauge transformation, it may be chosen as,
\begin{equation}
\label{eq:vacuum}
    A_\mu=Z_\mu=w_\mu=0,\qquad\phi=1,\qquad {\rm g}_{\mu\nu}=\eta_{\mu\nu},
\end{equation}
where $\eta_{\mu\nu}$ is the Minkowski metric. Linearizing the field equations \eqref{eq:A}-\eqref{eq:phi} and \eqref{eq:Ein} around this vacuum, one can read off the masses of the $Z$, $W$ and Higgs fields,
\begin{equation}
    m_{\rm Z}=\frac{1}{\sqrt{2}},\qquad m_{\rm W}=g\times m_{\rm Z},\qquad m_{\rm H}=\sqrt{\beta}\times m_{\rm Z},
\end{equation}
while the photon and graviton remain massless. These are the dimensionless masses in units of the dimensionful $Z$-boson mass ${\bm m}_{\rm Z}$, which thus sets the natural mass scale of the system (up to a factor $1/\sqrt{2})$. 

Throughout this work, the dimensionless parameters $g$, $g'$ and $\beta$ will be fixed to their physical values obtained by taking ratios of the different boson masses, and imposing the condition,
\begin{equation}
    g^2+g'^2=1,\qquad g\equiv\cos\theta_{\rm W},\qquad g'\equiv\sin\theta_{\rm W},
    \label{eq:weak_angle}
\end{equation}
where $\theta_{\rm W}$ is the weak mixing angle. Using the values $\bm m_{\rm W}\bm c^2\approx 80.37\,$GeV, $\bm m_{\rm Z}\bm c^2\approx91.19\,$GeV and $\bm m_{\rm H}\bm c^2\approx 125.2\,$GeV from Refs.\cite{Mohr:2024kco,PhysRevD.110.030001}, one has,
\begin{equation}
    \sin^2\theta_{\rm W}=1-\left(\frac{\bm m_{\rm W}}{\bm m_{\rm Z}}\right)^2\approx 0.223,\quad\quad\beta=\left(\frac{\bm m_{\rm H}}{\bm m_{\rm Z}}\right)^2\approx 1.88.
    \label{eq:ew_param_values}
\end{equation}

In principle, the dimensionless gravitational coupling \eqref{eq:kappa} can also be written in terms of the $Z$-boson mass,
\begin{equation}
    \kappa=\frac{4e^2}{\alpha}\left(\frac{\bm m_{\rm Z}}{\bm m_{\rm Pl}}\right)^2\,,\qquad\text{with}\quad {\bm m_{\rm Pl}}=\sqrt{\dfrac{\bm \hbar \bm c}{\bm G}}\,.
    \label{kappa_alpha}
\end{equation}
Choosing the physical value of the fine structure constant, $\alpha\approx 1/137$, and of the boson masses, one has $\bm m_{\rm Z}/\bm m_{\rm Pl}\sim 10^{-18}$ implying that $\kappa$ would be extremely small, of order $10^{-33}$. This reflects the large hierarchy between the Planck scale (set by $\bm m_{\rm Pl}$) and the electroweak scale (set by $\bm m_{\rm Z}$): gravitational effects are expected to be negligible in the context of the Standard Model\footnote{One remarkable exception is that of magnetic monopoles, whose energy diverges in flat space due to the $\rm{U}(1)_{\rm Y}$ part of the gauge group. In this case, coupling to gravity regularizes the monopole energy by concealing the central singularity behind an event horizon~\cite{Gervalle2023,Gervalle2024,Gervalle2025}.}. 
However, in this work, we adopt a different perspective. Keeping Newton's constant $\bm G$ at its physical value, we postulate the existence of a \textit{dark} electroweak sector governed by the same symmetries and boson mass ratios as in the Standard Model, but with a radically different overall energy scale (see Sec.~\ref{sec:numerical_physical} for details).

Such a \textit{dark} electroweak sector is not merely a theoretical convenience: strongly gravitating solitons in dark sectors of known models have attracted considerable interest in the context of dark matter and exotic compact objects. In particular, recent analyses of gravitational-wave events detected by LIGO-Virgo suggest that some signals, \emph{e.g.} GW190521, can be consistently interpreted as
the merger of two horizonless Proca stars composed of an ultralight vector
boson with mass $\sim 8.7\times 10^{-13}\,{\rm eV}/\bm c^2$
\cite{CalderonBustillo:2020fyi,CalderonBustillo:2022cja}. The model considered here promotes this scenario to a complete field-theoretic setting in which the vector mass originates from a Higgs mechanism rather than from an explicit Proca term, thereby offering a rich and self-consistent framework for the study of these objects from first principles.

Specifically, we will explore the existence of new horizonless compact objects -- akin to Proca stars -- in the gravitating \textit{dark} electroweak theory described by the action \eqref{eq:action-dimensionless}. The mass ratios of the gauge and Higgs bosons are fixed by the parameters $g$ and $\beta$, chosen at their Standard Model values, see Eqs.~\eqref{eq:weak_angle} and \eqref{eq:ew_param_values}, while the overall mass scale is assumed to be such that $\bm m_\text{W},\,\bm m_\text{Z},\,\bm m_\text{H} \sim 10^{-13}\,{\rm eV}/\bm c^2$. Newton's constant $\bm G$ is held at its physical value, corresponding to the Planck mass $\bm m_{\rm Pl}\sim 10^{19}\,{\rm GeV}/\bm c^2$.


\section{Ansatz, physical quantities and numerical approach}\label{sec:ansatz}

\subsection{Spherically symmetric fields}
\label{sec:ansatz_and_bc}

We now restrict to asymptotically flat, spherically symmetric configurations. Adopting isotropic coordinates, we parametrize the line element as,
\begin{equation}
    ds^2=-e^{2F_0(r)}dt^2+e^{2F_1(r)}\left(dr^2+r^2d\theta^2+r^2\sin^2\theta\,d\varphi^2\right).
\label{eq:metric-spherical-ansatz}
\end{equation}

The flat spacetime limit is recovered by setting $F_0=F_1=0$. In the matter sector, we assume a purely electric ansatz for the gauge fields, with a harmonic time dependence for the complex vector field, 
\begin{equation}
    A=A_t(r)\,dt,\qquad Z=Z_t(r)\,dt,\qquad w=e^{-i\omega t}\left[w_t(r)\,dt+i\,w_r(r)\,dr\right],
\label{eq:matter-spherical-ansatz}
\end{equation}
while for the Higgs,
\begin{equation}
    \phi=\phi(r).
\label{eq:higgs-spherical-ansatz}
\end{equation}

All radial functions are taken to be real, with the explicit factor $i$ in the radial component of $w_\mu$ introduced for convenience, as it makes the reduced field equations explicitly real. Notice that, in contrast with the pure Proca model \cite{Brito:2015pxa}, the complex field frequency $\omega$ can be eliminated from $w_\mu$ and absorbed into the gauge potential $A_\mu$ via the local gauge symmetry $\eqref{eq:residualU1}$. The harmonic time dependence of $w_\mu$ can therefore be seen as a gauge choice, though $\omega$ itself remains a meaningful parameter that we shall use to label the solutions. Without loss of generality, we take $\omega\geq 0$.

Substituting the ansatz \eqref{eq:metric-spherical-ansatz}-\eqref{eq:higgs-spherical-ansatz} into the field equations \eqref{eq:A}-\eqref{eq:phi}, \eqref{eq:Ein} yields a system of ordinary differential equations for the seven functions,
\begin{equation}
\left\{F_0(r),\,F_1(r),\,\phi(r),\,w_t(r),\,w_r(r),\,Z_t(r),\,A_t(r)\right\}.
\label{eq:reduced-functions}
\end{equation}

Together with the boundary conditions specified below, this system (not explicitly shown here) is solved numerically using two independent methods, see Sec.~\ref{app:num} for details.

Regularity at the origin implies that, around $r=0$, the metric functions $F_0$, $F_1$, the Higgs $\phi$, and the electric potentials $A_t$, $Z_t$, $w_t$ are even functions of $r$, whereas the radial component $w_r$ is odd. 
Accordingly, the local expansions near $r=0$ take the form, 
\begin{align}
F_0(r) &= F_{00}+F_{02}r^2+\mathcal{O}(r^4), &
F_1(r) &= F_{10}+F_{12}r^2+\mathcal{O}(r^4), \nonumber \\
\phi(r) &= \phi_0+\phi_2 r^2+\mathcal{O}(r^4), &
w_t(r) &= w_{t0}+w_{t2} r^2+\mathcal{O}(r^4), \nonumber \\
Z_t(r) &= Z_{t0}+Z_{t2} r^2+\mathcal{O}(r^4), &
A_t(r) &= A_{t0}+A_{t2} r^2+\mathcal{O}(r^4), \nonumber \\
w_r(r)&= w_{r1}r+\mathcal{O}(r^3),
\label{eq:origin-expansion-matter3}
\end{align}
where the quadratic coefficients $F_{02},\, F_{12},\,\phi_2,\,\dots$ are completely determined by the values of the fields at the origin, $F_{00},\,F_{10},\,\phi_0,\,\dots$, through the field equations. Similarly, the linear coefficient $w_{r1}$ is fixed algebraically by the Lorenz-type constraint \eqref{eq:omega-Lorenz} and is therefore not an independent parameter. The explicit expressions of these coefficients are,
\begin{subequations}
\label{eq:origin-coefficients}
\begin{align}
F_{02}
&=\frac{\kappa}{48}e^{-2F_{00}+2F_{10}}
\left[
4\phi_0^2\left(g^2w_{t0}^2+Z_{t0}^2\right)
-e^{2F_{00}}\beta\left(1-\phi_0^2\right)^2
\right],
\nonumber\\
F_{12}
&=-\frac{\kappa}{96}e^{-2F_{00}+2F_{10}}
\left[
2\phi_0^2\left(g^2w_{t0}^2+Z_{t0}^2\right)
+e^{2F_{00}}\beta\left(1-\phi_0^2\right)^2
\right],
\label{eq:origin-coefficients-metric}
\\[2mm]
\phi_2
&=-\frac{1}{24}e^{-2F_{00}+2F_{10}}
\left[
g^2w_{t0}^2+Z_{t0}^2
+e^{2F_{00}}\beta\left(1-\phi_0^2\right)
\right]\phi_0,
\label{eq:origin-coefficients-higgs}
\\[2mm]
w_{r1}
&=\frac{1}{3}e^{-2F_{00}+2F_{10}}
\left(gg'A_{t0}+{g'}^2Z_{t0}-\omega\right)w_{t0},
\nonumber\\
w_{t2}
&=\frac{1}{12}
\left[
e^{2F_{10}}g^2\phi_0^2w_{t0}
+6w_{r1}\left(g^2Z_{t0}-gg'A_{t0}+\omega\right)
\right],
\label{eq:origin-coefficients-W}
\\[2mm]
Z_{t2}
&=\frac{1}{12}
\left(e^{2F_{10}}\phi_0^2Z_{t0}-6g^2w_{t0}w_{r1}\right),
\nonumber\\
A_{t2}
&=\frac{gg'}{2}w_{t0}w_{r1}.
\label{eq:origin-coefficients-gauge}
\end{align}
\end{subequations}

The corresponding boundary conditions imposed at the origin in the numerical schemes are therefore,
\begin{equation}
    r=0:\quad F_0'=F_1'=\phi'=w_t'=A_t'=Z_t'=0,\quad w_r=0.
    \label{eq:bc_ori}
\end{equation}

At spatial infinity, asymptotic flatness and finite energy require the fields to approach the vacuum \eqref{eq:vacuum}. Linearizing the reduced field equations around this asymptotic vacuum yields the large-$r$ behavior,
\begin{align}
F_0(r)
&=-\frac{M}{r}+{\cal O}(r^{-2}),
&
F_1(r)
&=\frac{M}{r}+{\cal O}(r^{-2}),
\nonumber\\
A_t(r)
&=A_t(\infty)+\frac{Q_e}{r}+{\cal O}(r^{-2}),
&
Z_t(r)
&=c_Z\frac{e^{-m_{\rm Z}r}}{r}+\dots,
\nonumber\\
\phi(r)
&=1+c_H\frac{e^{-m_{\rm H}r}}{r}+\dots,
&
w_t(r)
&=c_W\frac{e^{-kr}}{r}+\dots,
\nonumber\\
w_r(r)
&=\frac{c_W\,\omega}{k^2}
\frac{(kr+1)e^{-kr}}{r^2}+\dots.
\label{eq:asymp_behav}
\end{align}
where $M$, $Q_e$, $c_Z$, $c_H$ and $c_W$ are real coefficients, and we have introduced the effective mass,
\begin{equation}
    k\equiv\sqrt{m_{\rm W}^2-\left[\omega-e A_t(\infty)\right]^2}.
    \label{eq:eff_mass}
\end{equation}

However, there exists a residual gauge freedom in Eq.~\eqref{eq:residualU1} of the form $\lambda\rightarrow\lambda_0+c t$,  with $\lambda_0$ and $c$ constants. The time-dependent part of this freedom is fixed by imposing $A_t(\infty)=0$, while $\lambda_0$ only induces an irrelevant constant phase rotation of $w_\mu$. Hence, any localized, finite-energy solution satisfies the bound-state condition,
\begin{equation}
    \omega< m_{\rm W}=\frac{g}{\sqrt{2}},
\end{equation}
which ensures the exponential decay of the complex $W$ field at infinity. The $1/r$ term in $A_t$ is the Coulomb tail of the massless electromagnetic field and therefore the coefficient $Q_e$ can be interpreted as the electric charge of the configuration, which can also be computed using an integral, as explained below. Similarly, $M$ coincides with the ADM mass of the solution.

From these asymptotics, we impose the following boundary conditions at infinity,
\begin{equation}
    r\to\infty :\quad F_0=F_1=w_t=w_r=A_t=Z_t=0,\quad \phi=1.
    \label{eq:bc_inf}
\end{equation}

Notice that the condition $A_t(r\to\infty)=0$ is an electromagnetic gauge choice, and $F_0(r\to\infty)=0$ fixes the normalization of the time coordinate.

\subsection{Quantities of interest}
\label{subsec:phys_quant}

We now turn to the definition of the physical quantities of interest, starting with the ADM mass. For the static, horizonless configurations considered here, it can be computed using the Komar volume integral \cite{Wald1984},
\begin{equation}
M=-\frac{\kappa}{8\pi}\int_\Sigma d^3x\,\sqrt{-{\rm g}}\,
\left(2\tensor{T}{^0_0}-T\right),
\label{eq:mass_integral}
\end{equation}
where $T=\tensor{T}{^\mu_\mu}$ is the trace of the stress-energy tensor and $\Sigma$ is a spacelike hypersurface with $t=\mathrm{const.}$ Alternatively, the mass can be read off from the asymptotic behavior of the metric,
\begin{equation}
-g_{tt}=1-\frac{2M}{r}+\mathcal{O}(r^{-2}).
\label{eq:mass_asymp}
\end{equation}

Comparing the mass computed from Eqs.~\eqref{eq:mass_integral} and \eqref{eq:mass_asymp} provides a test of the numerical accuracy of the solutions. We also introduce the rescaled mass,
\begin{equation}
\mathcal{M}=\frac{8\pi}{\kappa}M,
\end{equation}
which is related to the dimensionful mass through
$\bm{M}/\bm{m}_0=\mathcal{M}\times e^2/(4\pi\alpha)$, with
$\bm{m}_0=\sqrt{2}\,\bm{m}_{\rm Z}$. The origin of the prefactor
$e^2/(4\pi\alpha)$ will be explained in
Sec.~\ref{sec:numerical_physical}.

Similarly, the electric charge can be computed either from a volume integral of the charge density,
\begin{equation}
Q_e=-\frac{gg'}{4\pi}\int_\Sigma d^3x\,\sqrt{-{\rm g}}\,J^0,
\label{eq:charge_integral}
\end{equation}
or from the asymptotic behavior of the electric potential, as given in Eq.~\eqref{eq:asymp_behav},
\begin{equation}
A_t(r)=\frac{Q_e}{r}+\mathcal{O}(r^{-2}).
\label{eq:charge_asymp}
\end{equation}

Comparing the charge computed in both ways provides another estimate of the numerical accuracy. We also define the charge enclosed within a radius $r$ through the radial electric flux across a sphere $\sigma$ of constant radius,
\begin{equation}
q_e(r)
=-\frac{1}{4\pi}\int_\sigma d\theta\,d\varphi\,
\sqrt{-{\rm g}}\,F^{0r}
=-r^2e^{F_1-F_0}A_t'(r).
\label{eq:charge_function}
\end{equation}

Using Eq.~\eqref{eq:charge_asymp}, one recovers the total electric charge as the asymptotic value of $q_e$,
\begin{equation}
q_e(r\rightarrow\infty)=Q_e.
\end{equation}

We now turn to the compactness $\mathcal{C}$, a key diagnostic for assessing how closely our solutions can mimic a black hole. For objects without a hard surface, such as bosonic stars, no unique definition exists -- the freedom lying in the definition of an effective radius. A common choice is,
\begin{equation}
\mathcal{C}=\frac{M}{R_{99}},
\label{eq:compactness}
\end{equation}
where $R_{99}$ is an effective \textit{areal} radius enclosing $99\%$ of the total mass $M$, defined through a suitable mass function. For example, one could truncate the mass integral \eqref{eq:mass_integral} at a finite radius. However, our configurations contain the Coulomb tail of the electric field, whose contribution to the stress-energy tensor decays as a power law. Such a definition would therefore lead to artificially large values of $R_{99}$.

We instead introduce the areal, Schwarzschild-like coordinate
$\tilde r=r\,e^{F_1(r)}$, assuming that the map $r\mapsto\tilde r$ is monotonic, and parametrize the metric as
\begin{align}
ds^2={}&-\zeta(\tilde r)
\left(
1-\frac{2m(\tilde r)}{\tilde r}
+\frac{q(\tilde r)^2}{\tilde r^2}
\right)dt^2
+\left(
1-\frac{2m(\tilde r)}{\tilde r}
+\frac{q(\tilde r)^2}{\tilde r^2}
\right)^{-1}d\tilde r^2
\notag\\
&\quad
+\tilde r^2\left(d\theta^2+\sin^2\theta\,d\varphi^2\right),\label{eq:met_sch_like}
\end{align}
where
$q(\tilde r)=\sqrt{\kappa/2}\times q_e(\tilde r)$ is the rescaled charge enclosed within the radius $\tilde r$, $\zeta(\tilde r)$ is a new metric function, and $m(\tilde r)$ is the mass function. Comparing this metric with Eq.~\eqref{eq:metric-spherical-ansatz} and using Eq.~\eqref{eq:charge_function}, we obtain, as a function of the isotropic coordinate used in our numerical calculations,
\begin{equation}
m(r)=\frac{1}{2}r\,e^{F_1}
\left[
1-\left(1+r\,F_1'\right)^2
+\frac{\kappa}{2}r^2e^{-2F_0}\left(A_t'\right)^2
\right].
\label{eq:mass_function}
\end{equation}

In an exact Reissner-Nordstr\"om (RN) exterior, this definition gives
$m(r)=M$ identically, confirming that the explicit charge term introduced in Eq.~\eqref{eq:met_sch_like} effectively subtracts the Coulomb contribution to the mass function. For the solutions considered here, the remaining fields in the exterior are massive, hence $m(r)$ approaches $M$ exponentially fast. We then determine the isotropic radius $r_{99}$ through the equation,
\begin{equation*}
m(r_{99})=0.99\times M,
\end{equation*}
which we have verified admits a single root for all numerical solutions considered here. As a consistency check, we also find that $q_e(r_{99})\simeq Q_e$, showing that this radius encloses nearly all of the electric charge. The corresponding areal radius entering Eq.~\eqref{eq:compactness} is then,
\begin{equation}
R_{99}=r_{99}e^{F_1(r_{99})}.
\end{equation}

Finally, we introduce the mass-to-charge ratio,
\begin{equation}
\Upsilon
=\sqrt{\frac{2}{\kappa}}\frac{M}{|Q_e|}
=\frac{\sqrt{2\kappa}}{8\pi}
\frac{\mathcal{M}}{|Q_e|}.\label{eq:Upsilon}
\end{equation}

For a RN black hole, the condition for the existence of an event horizon is $\Upsilon\geq1$, with equality in the extremal case. As we shall see, our horizonless solutions can approach the extremal value $\Upsilon=1$ in certain regions of the parameter space.

\subsection{Numerical approaches}\label{app:num}

 The coupled nonlinear ordinary differential equations for the seven radial functions $\{F_0,F_1,w_t,w_r,\phi,Z_t,A_t\}$ are solved numerically as a boundary-value problem. We compactify the semi-infinite interval $r\in[0,\infty)$ according to,
\begin{equation}
    x=\frac{r}{c+r},
    \qquad
    x\in[0,1],
\end{equation}
where $x$ is the compactified radial coordinate and $c$ is an input parameter that controls the distribution of radial points. The boundary conditions \eqref{eq:bc_ori} and \eqref{eq:bc_inf} are imposed directly at $x=0$ and $x=1$. We emphasize that the expansions discussed in Sec.~\ref{sec:ansatz_and_bc} justify the regularity conditions at the origin, but neither solver relies on a truncated series expansion.

We use two independent numerical implementations, based respectively on the finite-difference code CADSOL and the finite-element code FreeFEM. Numerical accuracy is measured using several independent diagnostics: the residuals of the field equations, the Lorenz-like constraints, the agreement between the Komar and asymptotic definitions of the mass, and the agreement between the volume and flux definitions of the charge. 
In addition, since the unitary-gauge form of the electroweak theory \eqref{eq:Lagr} is ill-defined at zeroes of the Higgs field \cite{Rubakov:2002fi}, we explicitly monitor $\phi$ throughout the numerical construction and verify that it remains strictly positive for all solutions presented below.

The results presented in this work have been obtained using an equidistant grid of $400$ points in CADSOL and a non-equidistant grid of $1000$ points in FreeFEM. In the latter case, the resolution is refined near the asymptotic boundary $x=1$, where small variations in $x$ correspond to increasingly large variations in $r$. Typical relative errors are of order $10^{-10}$ with FreeFEM and $10^{-6}$ with CADSOL. Further details on these numerical methods can be found in Refs.~\cite{SCHONAUER2001473,schmauder1992cadsol,SCHONAUER1989279,MR3043640,Gervalle:2022fze,Herdeiro:2025blx,Gervalle:2024poh}; see also our forthcoming work~\cite{BSFactory}.

Both methods rely on a Newton-Raphson relaxation and, therefore, require a sufficiently accurate initial guess. We construct the first solutions using the radial profiles of Proca-Higgs stars as initial seeds~\cite{herdeiro2023procahiggs,Brito:2024biy}. 
Once a converged solution is obtained, the remaining configurations are constructed by continuation, using the previous solution as an initial guess while varying one of the input parameters by a small amount.

The numerical solutions are specified by four independent input parameters, namely
\(\kappa\), \(\beta\), \(\omega\), and \(\theta_{\rm W}\). The parameter \(\kappa\) controls the strength of the gravitational coupling, while \(\beta\) fixes the Higgs self-coupling. The weak mixing angle \(\theta_{\rm W}\) fixes the relative strength of the two gauge couplings through \(g=\cos\theta_{\rm W}\) and \(g'=\sin\theta_{\rm W}\). Finally, the frequency \(\omega\) determines the harmonic time dependence of the charged vector field $w_\mu$ and parametrizes the different families of solutions. Once these dimensionless parameters are fixed, the corresponding physical configurations are obtained by choosing the overall mass scale (for example, the dimensionful value of the $W$-boson mass, see Sec.~\ref{sec:numerical_physical}).

\section{The solutions}\label{Sec:solutions}

We now present the numerical solutions, which consist in families of everywhere regular, asymptotically flat field configurations. Throughout this section, we fix the Higgs coupling $\beta$ and weak mixing angle $\theta_{\rm W}$ to their Standard Model values, given in Eq.~\eqref{eq:ew_param_values}, and vary the gravitational coupling $\kappa$ and frequency $\omega$.

\begin{figure}[h!]
    \makebox[\linewidth][c]{
        \begin{subfigure}[b]{0.485\textwidth}
        \centering
            \includegraphics[width=\linewidth]{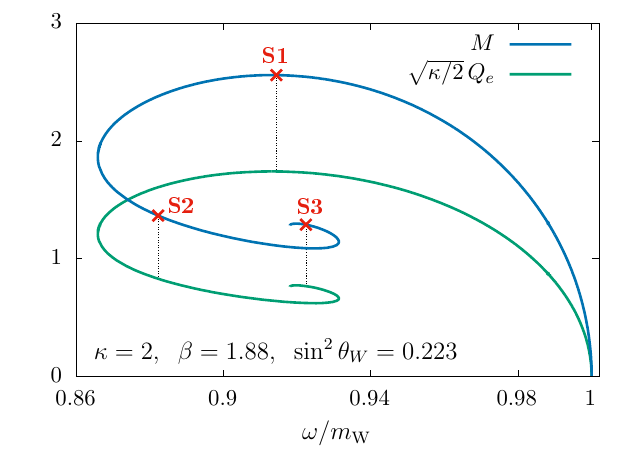}
        \end{subfigure}
        \qquad{}
        \begin{subfigure}[b]{0.485\textwidth}
        \centering
            \includegraphics[width=\linewidth]{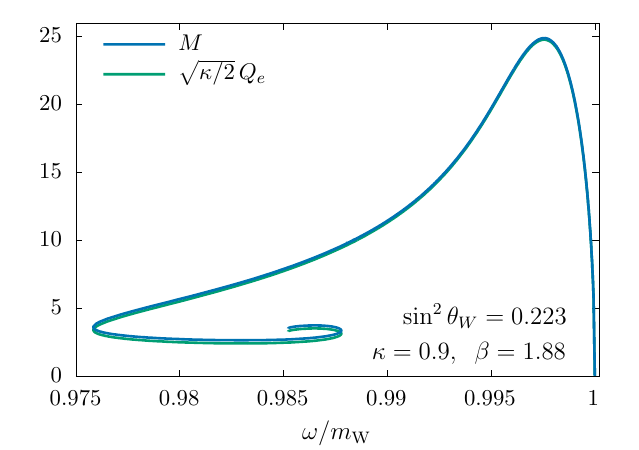}
        \end{subfigure}
    }

\caption{The ADM mass $M$ and the rescaled electric charge
$\sqrt{\kappa/2}\times Q_e$ as functions of $\omega/m_{\rm W}$ for the
families with $\kappa=2$ (left) and $\kappa=0.9$ (right). The red
crosses in the left panel mark the representative solutions S1,
S2, and S3, whose profiles are shown in
Figs.~\ref{sol1_metric}-~\ref{sol1}.}
    \label{fig_seq_MQ}
\end{figure}

The asymptotic behavior \eqref{eq:asymp_behav} implies that localized configurations exist for $k^2>0$, equivalently for $0\leq\omega< m_{\rm W}$. At a fixed $\kappa$, in the limit $k\approx0$ where $\omega\approx m_{\rm W}$, one recovers the usual Newtonian branch, where the configurations are very diluted and weakly gravitating, with $M,Q_e\to0$. As one decreases $\omega$ and moves away from this limit, the solutions become more compact and their mass increases until a maximum is reached. After this point, the mass starts decreasing and one reaches a minimal frequency $\omega_{\rm min}$ whose value depends on $\kappa$. Further continuation of the family of solutions leads to the standard spiral structure in the parameter space, as commonly observed for scalar boson stars and Proca(-Higgs) stars \cite{Liebling:2012fv,Brito:2015pxa,herdeiro2023procahiggs}. 
These aspects are illustrated in Fig.~\ref{fig_seq_MQ} for the families with $\kappa=2$ and $\kappa=0.9$.

\begin{figure}[h!]
    \makebox[\linewidth][c]{
        \begin{subfigure}[b]{0.485\textwidth}
        \centering
            \includegraphics[width=\linewidth]{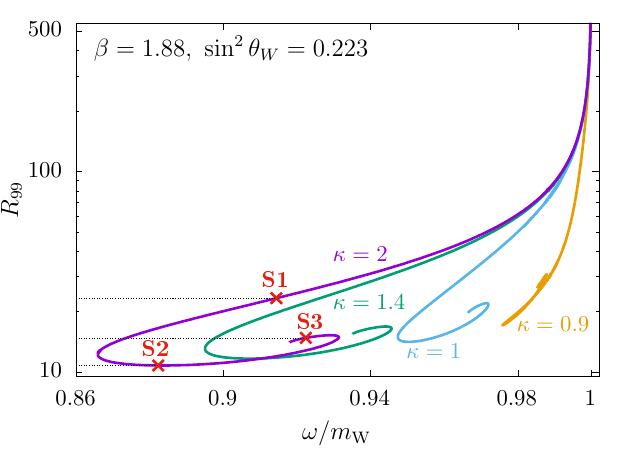}
        \end{subfigure}
        \qquad{}
        \begin{subfigure}[b]{0.485\textwidth}
        \centering
            \includegraphics[width=\linewidth]{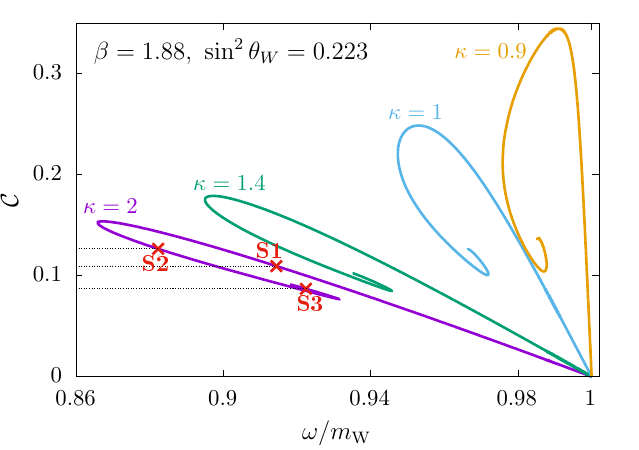}
        \end{subfigure}
    }

\caption{The effective radius $R_{99}$ (left) and compactness
$\mathcal{C}$ (right) as functions of $\omega/m_{\rm W}$ for the
families with $\kappa\in\{2,\,1.4,\,1,\,0.9\}$. The red crosses mark
the representative solutions S1, S2 and S3 on the
$\kappa=2$ family.}
    \label{fig_R99_C}
\end{figure}

\begin{figure}[h!]
    \makebox[\linewidth][c]{
        \begin{subfigure}[b]{0.485\textwidth}
        \centering
            \includegraphics[width=\linewidth]{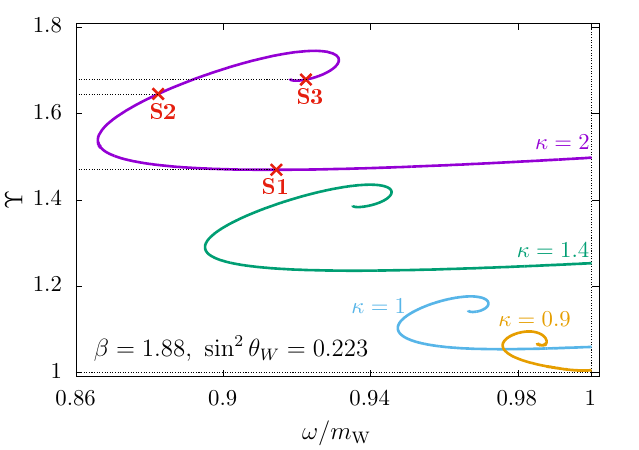}
        \end{subfigure}
        \qquad{}
        \begin{subfigure}[b]{0.485\textwidth}
        \centering
            \includegraphics[width=\linewidth]{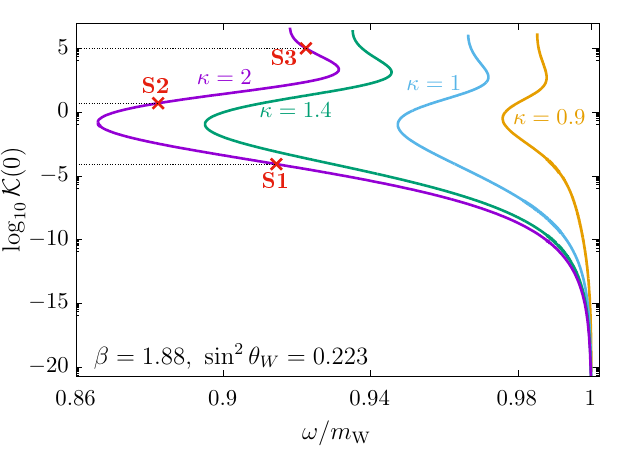}
        \end{subfigure}
    }

    \caption{The mass-to-charge ratio $\Upsilon$ (left) and the logarithm
of the central Kretschmann scalar, $\log_{10}\mathcal{K}(0)$ (right),
as functions of $\omega/m_{\rm W}$ for the families with
$\kappa\in\{2,\,1.4,\,1,\,0.9\}$. The red crosses mark the
representative solutions S1, S2, and S3 on the $\kappa=2$
family.}
    \label{fig_MoQ_K}
\end{figure}

\begin{figure}[h!]
	\makebox[\linewidth][c]{%
		\begin{subfigure}[b]{0.485\textwidth}
			\centering
\includegraphics[width=\linewidth]{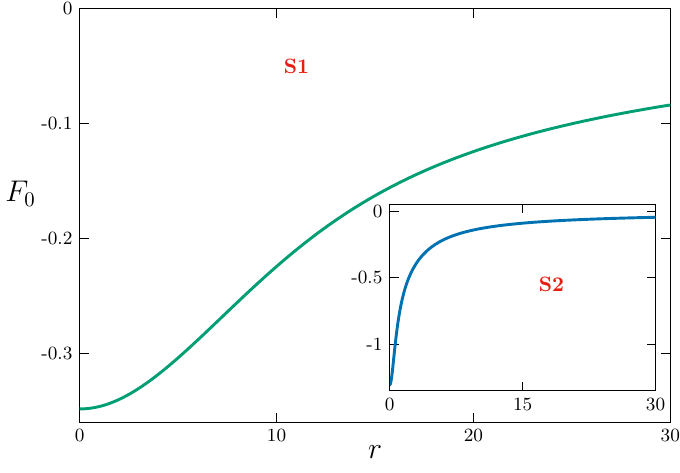}
		\end{subfigure}%
        \qquad{}
		\begin{subfigure}[b]{0.485\textwidth}
			\centering
\includegraphics[width=\linewidth]{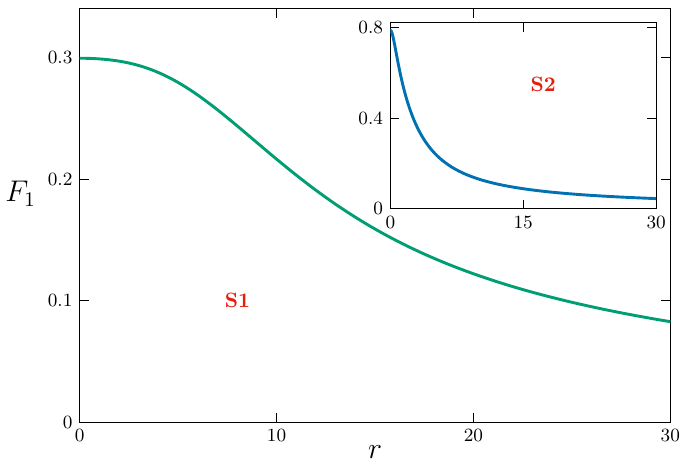}
		\end{subfigure}%
        }
          \\
  \\
 \makebox[\linewidth][c]{%
		\begin{subfigure}[b]{0.485\textwidth}
			\centering
\includegraphics[width=\linewidth]{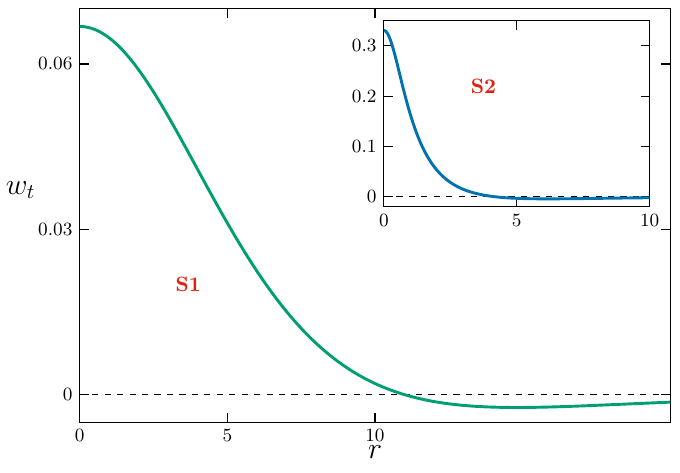}
		\end{subfigure}%
        \qquad{}
		\begin{subfigure}[b]{0.485\textwidth}
			\centering
\includegraphics[width=\linewidth]{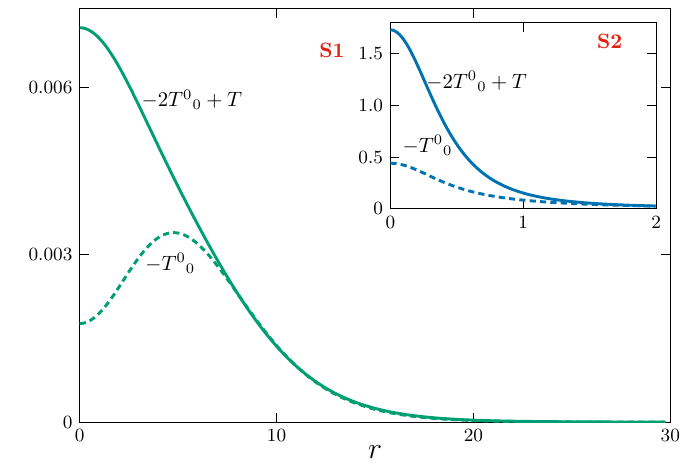}
		\end{subfigure}%
	} 
\caption{The metric functions $F_0$ (top left) and $F_1$ (top right) for the
representative solution S1. The insets show the corresponding
profiles for S2. The $W$ field component $w_t$ (bottom left), and the Komar mass density $-2T^0_{\ 0}+T$ and energy density $-T^0_{\ 0}$ (bottom right), for the representative solution S1. The insets show the corresponding profiles for S2. For all the solutions computed, $w_t$ has one node. The Komar mass density reaches its maximum at the center of the star, whereas the energy density may, in general, reach its maximum away from the center.}
\label{sol1_metric}
\end{figure}

\begin{figure}[h!]
 \makebox[\linewidth][c]{%
		\begin{subfigure}[b]{0.485\textwidth}
			\centering
\includegraphics[width=\linewidth]{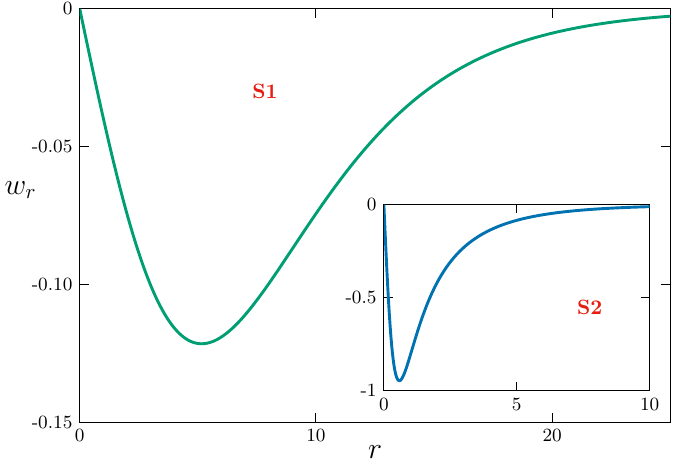}
		\end{subfigure}%
        \qquad{}
		\begin{subfigure}[b]{0.485\textwidth}
			\centering
\includegraphics[width=\linewidth]{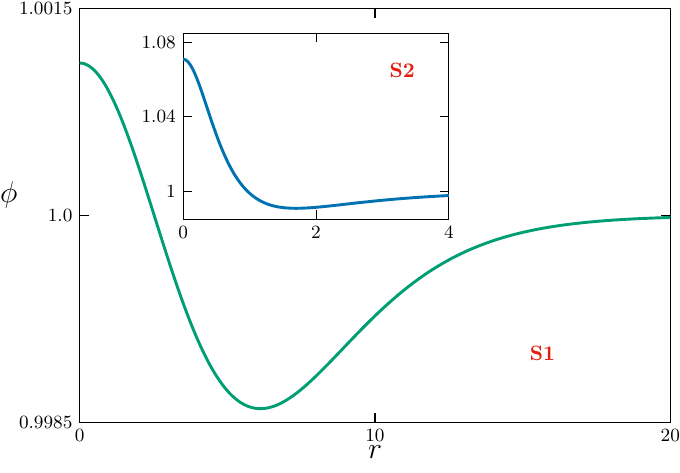}
		\end{subfigure}%
	}
  \\
  \\
 \makebox[\linewidth][c]{%
		\begin{subfigure}[b]{0.485\textwidth}
			\centering
\includegraphics[width=\linewidth]{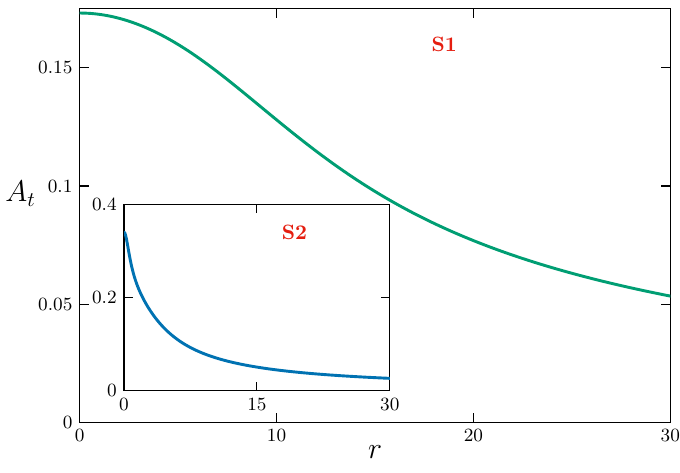}
		\end{subfigure}%
        \qquad{}
		\begin{subfigure}[b]{0.485\textwidth}
			\centering
\includegraphics[width=\linewidth]{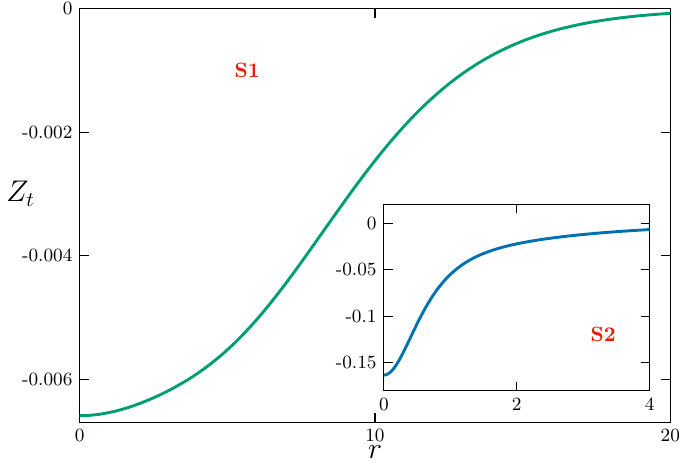}
		\end{subfigure}%
	} 
	\caption{
	{\small
The $W$-field component $w_r$ (top left), Higgs field $\phi$ (top right), electric potential $A_t$ (bottom left) and $Z$-boson potential $Z_t$ (bottom right) for the representative solution S1. The insets show the corresponding profiles for S2.}
		\label{sol1}
  }
\end{figure}

For $\kappa=2$, we select three representative solutions S1, S2 and S3, indicated by red crosses in Figs.~\ref{fig_seq_MQ}-\ref{fig_MoQ_K}, and report their physical properties in Table~\ref{tab_sols}. The field profiles of S1 and S2 are shown in Figs.~\ref{sol1_metric}--\ref{sol1}. Interestingly, the Higgs amplitude $\phi$, shown in the top-right panel of Fig.~\ref{sol1}, is non-monotonic: it exceeds the vev in the central region and falls below the vev in the exterior. The bottom panels of Fig.~\ref{sol1} show that the electric and $Z$ potentials have opposite signs. Consequently, the corresponding electric fields, $E_i=F_{0i}$ and $E^{(Z)}_i=Z_{0i}$, are oppositely directed. This follows directly from the source terms in the right-hand sides of Eqs.~\eqref{eq:A} and \eqref{eq:Z}, which enter with opposite signs. In these bosonic stars, the electric field is generated by the condensate of charged $W$ bosons inside the configuration.

In Fig.~\ref{fig_R99_C}, we show the effective radius $R_{99}$ and compactness $\mathcal{C}$ for representative values of $\kappa$. Along the Newtonian branch, $R_{99}$ grows arbitrarily large while $\mathcal{C}$ approaches zero as the configurations become increasingly dilute, converging towards the trivial vacuum in the limit $\omega\to m_{\rm W}$. Moving away from the Newtonian branch, $R_{99}$ decreases down to values $\sim 10$, while $\mathcal{C}$ grows and reaches a maximum before declining close to $\omega_{\rm min}$. Both quantities subsequently exhibit the characteristic spiraling behavior. The figure also shows that $\omega_{\rm min}(\kappa)$ is an increasing function of $\kappa$, so that the accessible $\omega$-range becomes progressively narrower for smaller $\kappa$. At the same time, the maximal compactness grows and one has,
\begin{equation}\label{eq:max-compactnesses}
 {\cal C}_{\rm max}\simeq
  0.154,\ 0.179,\ 0.249,\ 0.344
  \qquad\hbox{for}\qquad
  \kappa=2,\ 1.4,\ 1,\ 0.9.
\end{equation}
This trend results from the growth of the maximal mass as $\kappa$ decreases (see Fig.~\ref{fig_seq_MQ}), while the effective radius remains of the same order of magnitude. However, none of the families displayed here reaches ${\cal C}=0.5$, corresponding to the Schwarzschild value. For comparison, a RN black hole with the same mass $M$ and electric charge $Q_e$ would have,
\begin{equation}
   r_+=M\left(1+\sqrt{1-\Upsilon^{-2}}\right)
   \quad\Rightarrow\quad
   \mathcal{C}=\frac{M}{r_+}=\frac{1}{1+\sqrt{1-\Upsilon^{-2}}},
\label{eq:RN-comparison}
\end{equation}
where $r_{+}$ denotes the outer horizon radius. Its compactness therefore ranges from $1/2$ in the Schwarzschild limit ($Q_e\to0$) to $1$ in the extremal limit.

Figure~\ref{fig_MoQ_K} shows the mass-to-charge ratio $\Upsilon$ and the central Kretschmann scalar value $\mathcal{K}(0)$ for our families of solutions. As seen in the left panel, $\Upsilon$ exhibits the spiral structure inherited from $M$ and $Q_e$, and approaches a finite value in the limit $\omega\to m_{\rm W}$. Overall, $\Upsilon$ remains of order unity and approaches $1$ for smaller values of $\kappa$, consistently with the behavior $M\approx\sqrt{\kappa/2}\times Q_e$ visible in the right panel of Fig.~\ref{fig_seq_MQ}. The right panel of Fig.~\ref{fig_MoQ_K} shows that $\mathcal{K}(0)$ grows along each family, indicating an increasingly strong spacetime curvature in the central region as one moves deeper into the $M(\omega)$ spiral. 

For a fixed $\omega\in[\omega_{\rm min},m_{\rm W})$, we attempted to continue the solutions numerically toward smaller values of the gravitational coupling $\kappa$. However, the continuation becomes increasingly difficult for $\kappa\lesssim 0.9$, and we were unable to reach the flat-space limit when $\beta$ and $\theta_{\rm W}$ are fixed at their Standard Model values\footnote{Nevertheless, we were able to construct flat-spacetime solutions for other choices of these parameters.}. This behavior can be heuristically understood by considering the Newtonian interaction between two $W$ bosons with the same charge. At separations much larger than their Compton wavelength, $r\gg1/m_{\rm W}$, only the gravitational and electromagnetic interactions remain relevant. In our dimensionless conventions, the corresponding interaction potential is,
\begin{equation}
V(r)
=-\frac{\kappa m_{\rm W}^2}{8\pi r}
+\frac{e^2}{4\pi r}
=\frac{e^2}{4\pi r}
\left(1-\frac{\kappa}{4g'^2}\right),
\label{eq:long-range-W-interaction}
\end{equation}
so that the two long-range forces balance at,
\begin{equation}
\kappa_c
=\frac{2e^2}{m_{\rm W}^2}
=4g'^2
=4\sin^2\theta_{\rm W}
\simeq0.892.
\label{eq:kappa-critical}
\end{equation}
According to this simple toy-model, for $\kappa>\kappa_c$, gravitational attraction dominates over electric repulsion, resulting in a net attractive long-range interaction between the two bosons. As their separation becomes comparable to their Compton wavelength, short-range interactions become relevant and prevent their collapse, thereby allowing a bound state to form. Conversely, for $\kappa<\kappa_c$, the long-range interaction is repulsive, as gravity is no longer sufficiently strong to compensate the electric repulsion, and the formation of a bound state is therefore not expected.  

It is important to stress that this heuristic model only characterizes the non-relativistic long-range force balance and the bound $\kappa>\kappa_c$ should therefore not be interpreted, by itself, as a general condition for the existence of bound-state solutions. As we shall see below, the numerical solutions actually cease to exist at slightly smaller $\kappa$. Nevertheless, the critical value $\kappa_c$ given in Eq.~\eqref{eq:kappa-critical} coincides with the encountered numerical difficulties. Moreover, this result allows us to understand the value of the mass-to-charge ratio $\Upsilon$ along the Newtonian branch and why it approaches unity as $\kappa$ decreases toward $\kappa_c$. In the limit $\omega\rightarrow m_{\rm W}$, we see from Eq.~\eqref{Eq:diffmass} that 
$\mathcal{M}\approx4\pi|Q_e|/(\sqrt{2}\,g')$. Substituting this into Eq.~\eqref{eq:Upsilon} yields,
\begin{equation}
\Upsilon\approx\frac{\sqrt{\kappa}}{2g'}=\sqrt{\frac{\kappa}{\kappa_c}}\,.
\label{eq:estim_ups}
\end{equation}

Hence, in the Newtonian branch, $\Upsilon$ approaches unity as $\kappa\to\kappa_c$, in agreement with behavior observed in the left panel of Fig.~\ref{fig_MoQ_K}. In addition, for $\kappa>\kappa_c$, the estimate \eqref{eq:estim_ups} remarkably coincides with the numerical values reached at the right border of the plot. This agreement suggests that the simple Newtonian model captures the essential physics of diluted configurations with $\omega\approx m_{\rm W}$, for which we recall that $R_{99}\gg 1/m_{\rm W}$.

\begin{figure}[h!]
    \makebox[\linewidth][c]{
        \begin{subfigure}[b]{0.485\textwidth}
        \centering
            \includegraphics[width=\linewidth]{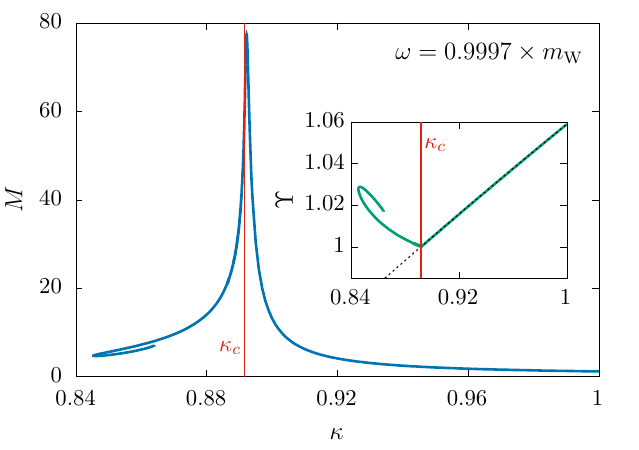}
        \end{subfigure}
        \qquad{}
        \begin{subfigure}[b]{0.485\textwidth}
        \centering
            \includegraphics[width=\linewidth]{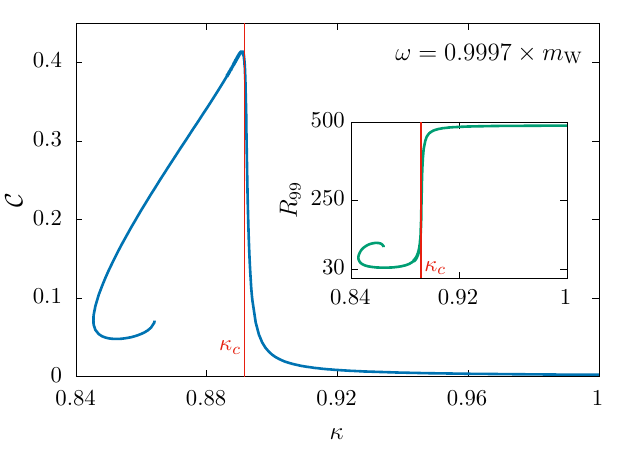}
        \end{subfigure}
    }
    \\[2ex]
    \makebox[\linewidth][c]{
        \begin{subfigure}[b]{0.485\textwidth}
        \centering
            \includegraphics[width=\linewidth]{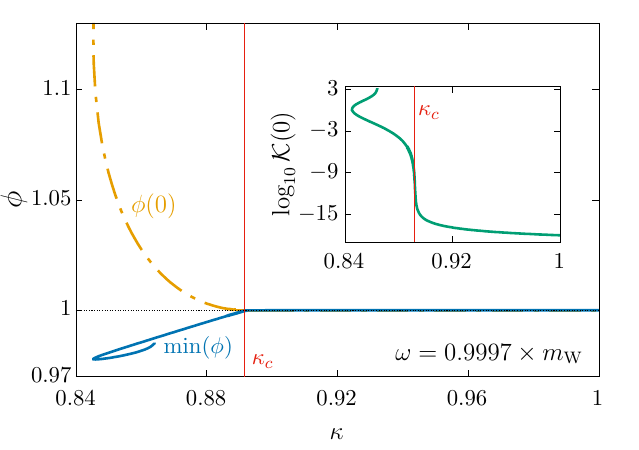}
        \end{subfigure}
        \qquad{}
        \begin{subfigure}[b]{0.485\textwidth}
        \centering
            \includegraphics[width=\linewidth]{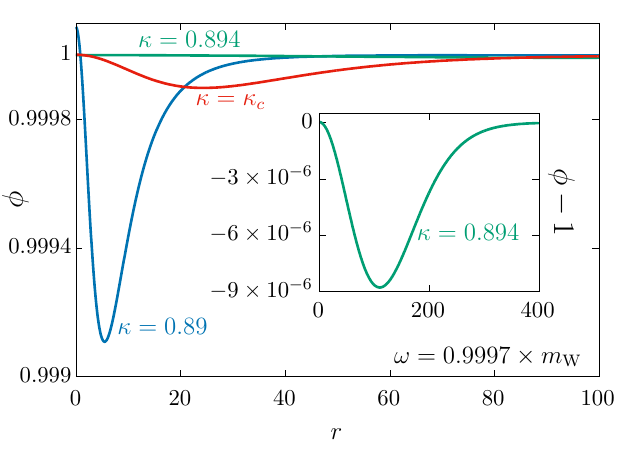}
        \end{subfigure}
    }

    \caption{Dependence of the solution properties on the gravitational coupling $\kappa$, for the fixed frequency $\omega=0.9997\times m_{\rm W}$. Top left: ADM mass $M$, with the mass-to-charge ratio $\Upsilon$ shown in the inset. The dashed curve represents the Newtonian estimate $\Upsilon\simeq\sqrt{\kappa/\kappa_c}$. Top right: compactness ${\cal C}$, with the effective radius $R_{99}$ shown in the inset. Bottom left: the central value $\phi(0)$ and minimum value $\min(\phi)$ of the Higgs field, with the central Kretschmann scalar ${\cal K}(0)$ shown in the inset.  The vertical red lines mark the critical coupling $\kappa_c=4g'^2=0.892$. Near $\kappa_c$, the mass and compactness reach their largest values while the effective radius decreases sharply. For $\kappa<\kappa_c$, the Higgs field departs substantially from the vev, its central value grows rapidly along with the central curvature, and the minimal value of the Higgs develops a back-bending followed by a spiral.
    Bottom right: Higgs profile $\phi(r)$ for three representative values of $\kappa\in\{0.89,0.892,0.894\}$ around $\kappa_c$.}
    \label{fig_varkap}
\end{figure}
To elucidate the role of the critical value $\kappa_c$, we present in Fig.~\ref{fig_varkap} the dependence of the physical quantities on $\kappa$ for a fixed frequency $\omega=0.9997\times m_{\rm W}$. In the top-left panel, the mass peaks at $\kappa\approx\kappa_c$, the maximum being attained slightly above the critical value, and decreases for values $\kappa<\kappa_c$ until a minimal value, $\kappa_{\rm min}$, of the gravitational coupling is reached. This point corresponds to a back-bending of the solution family, which can be continued toward larger values of $\kappa$. A second back-bending occurs shortly thereafter, indicating the onset of a spiral structure. The growth of the mass as $\kappa$ is lowered from above $\kappa_c$ can be understood from the increasing contribution of electromagnetic repulsion relative to gravitational attraction. As the latter becomes less effective, the (negative) binding energy is progressively reduced, thereby increasing the total mass. The inset shows that the mass-to-charge ratio satisfies $\Upsilon>1$ throughout, approaching the limiting value $\Upsilon=1$ from above as $\kappa\to\kappa_c$. We note that the unit value is attained precisely at $\kappa_c$, marking the transition between two regimes. For $\kappa>\kappa_c$, the behavior of $\Upsilon$ is remarkably well-described by the square-root law in Eq.~\eqref{eq:estim_ups}, represented by a dashed curve on the figure. For $\kappa<\kappa_c$, $\Upsilon$ exhibits the spiralling behavior and no longer follows the law \eqref{eq:estim_ups}.

The compactness $\mathcal{C}$ and effective radius $R_{99}$ are shown in the top-right panel of Fig.~\ref{fig_varkap}. Here, $\kappa_c$ marks a pronounced transition in the effective radius, which drops sharply from $R_{99}\sim 500$ to $R_{99}\sim 30$. Notably, $R_{99}$ is approximately constant above $\kappa_c$. Combined with the peak in the mass, this sharp decrease in size drives the compactness up to $\mathcal{C}\approx 0.4$ around $\kappa\approx\kappa_c$. The configurations are therefore no longer dilute at this point, despite having $\omega\approx m_{\rm W}$. This illustrates that the critical value is associated with a qualitative change in the structure of the bosonic stars. Above $\kappa_c$, the solutions are essentially ``Newtonian'' in the sense that they are well-described by the simple heuristic model discussed above. Close to $\kappa_c$, however, the solutions can no longer be supported by the long-range gravitational binding alone, as it is counterbalanced by electromagnetic repulsion. The configurations thus enter a new regime in which they are maintained by the nonlinear/short-range structure of the full theory. This mechanism allows the bosonic stars to persist even below $\kappa_c$. The subsequent spiral structure indicates that this nonlinear regime cannot be continuously extended to the flat-space limit: our numerical results show that the family remains confined to $\kappa\in[\kappa_{\rm min},\kappa_c]$, with $\kappa_{\rm min}$ marking a back-bending point rather than the endpoint of the family.

This transition from the Newtonian to nonlinear regime is further supported by the behavior of the Higgs field, shown in the bottom panels of Fig.~\ref{fig_varkap}. Above $\kappa_c$, the Higgs remains essentially frozen at its vev, $\phi\approx 1$, as the configurations are very dilute and close to the trivial vacuum. This is consistent with the electroweak sector effectively reducing to ordinary electromagnetism in this regime, where the long-range interactions dominate the dynamics. We emphasize that the Higgs is not exactly at the vev for $\kappa>\kappa_c$, but exhibits small yet nonzero deviations (see the inset in the bottom right panel of the figure). Below $\kappa_c$, the minimal value of the Higgs starts deviating linearly from the vev, while its central value $\phi(0)$ rapidly grows. The full electroweak sector therefore becomes increasingly nontrivial, with short-range interactions providing additional binding needed to maintain the bosonic stars. Beyond the back-bending point at $\kappa_{\rm min}$, the minimal value of $\phi$ starts spiraling, while $\phi(0)$ seems to grow arbitrarily large. Finally, the inset in the bottom left panel shows the central value of the Kretschmann scalar, $\mathcal{K}(0)$, which remains very small for $\kappa>\kappa_c$, characteristic of the weak-field regime, but grows rapidly for $\kappa<\kappa_c$, indicating that the solutions also enter a strongly curved, nonlinear gravitational regime.

 In the pure Proca model, the spherical configurations studied in Ref.~\cite{Brito:2015pxa} attain their maximum Komar mass density away from the origin, unlike ground-state scalar stars, whereas the dynamically preferred prolate configurations attain it at the center~\cite{Herdeiro:2023wqf}. In the present model, the maximum Komar mass density occurs at the center for all the computed solutions. On the other hand, along the branch expected to be stable, the maximum energy density occurs away from the star center (see Fig.~\ref{sol1_metric}). This distinct behavior, together with the additional electromagnetic, $Z$, and Higgs degrees of freedom, suggests that the dynamics of these solutions may differ substantially from those of pure Proca stars. Nevertheless, neither the location of the maximum energy density nor that of the maximum Komar mass density is, by itself, a stability criterion. Linear and nonlinear stability analyses are still necessary to determine which configurations are dynamically stable.

\begin{table}[ht]
    \centering
\begin{tabular}{l|c c c c c}
 \toprule
 & $\omega$ & $M$ & $Q_e$ & $\mathcal{C}$ & $\mathcal{K}(0)$ \\
 \hline
 S1 & $0.57$  & $2.558$ & 1.740 & $0.109$ & $8.309\times 10^{-5}$ \\
 S2 & 0.55  & 1.366 & 0.830 &  0.127 & $5.025$ \\
 S3 & 0.575 & 1.289   & 0.768  & 0.087 & $1.028\times 10^{+5}$ \\
 \bottomrule
\end{tabular}
\caption{Properties of the representative solutions S1, S2, and S3, corresponding, respectively, to the maximal-mass point, the returning branch, and the inner spiral of the $M(\omega)$ diagram. We display the complex field frequency $\omega$, ADM mass $M$, electric charge $Q_e$, compactness $\mathcal{C}$, and central Kretschmann scalar $\mathcal{K}(0)$. Despite having masses of the same order, S3 exhibits an increase of approximately nine orders of magnitude in $\mathcal{K}(0)$ as compared to S1. 
}
\label{tab_sols}
\end{table}

Although we do not show the corresponding plot, we mention that increasing the Higgs self-coupling $\beta$ makes the potential around the vacuum steeper, 
thereby increasing the energetic cost of departures from $\phi=1$. The role of the proper Higgs vev ${\bm\Phi}_0$ is indirect through $\kappa$, see Eq.~\eqref{eq:kappa}, as we are working in the dimensionless scheme. Increasing the vev, or equivalently $\kappa$, likewise makes it energetically costly for the Higgs field to depart from its vev. Conversely, smaller values of $\beta$ or $\kappa$ allow the Higgs to deviate more substantially from $\phi=1$ and hence play a more active role in the configurations, see the bottom panels of Fig.~\ref{fig_varkap} where $\kappa$ is varied and, for instance, the discussion in \cite{herdeiro2023procahiggs,Herdeiro:2024pmv}. We have checked that decreasing $\beta$ also leads to larger deviations from the Higgs vev. At the same time, even for the reference solutions with $\beta=1.88$ and $\kappa=2$, these deviations can be substantial: while they remain below $0.15\%$ for S1, they reach $7\%$ for S2 and more than $30\%$ for S3. Thus, the chosen parameter values seem sufficient to have an active Higgs field along the solution family. 

It is also worth emphasizing that the existence of these non-topological solutions does not depend on the particular physical value of $\bm{m}_{\rm W}$, as long as $\bm{m}_{\rm W}\neq0$. Indeed, $\bm{m}_{\rm W}$ only fixes the physical scale used to convert the dimensionless numerical output into dimensionful quantities. Therefore, at the level of the dimensionless equations, the same family of solutions can be interpreted for any nonzero value of the physical $W$-boson mass. The requirement that it should be ultralight is not, therefore, a condition for the existence of the solutions themselves. Rather, it is a phenomenological requirement: since the dimensionful mass of these bosonic stars scale with the inverse of the boson mass, an ultralight $\bm{m}_{\rm W}$ leads to configurations with astrophysical masses and sizes, see Table \ref{tab:physical_scales}. For larger values of $\bm{m}_{\rm W}$, the same solutions would simply correspond to microscopic objects.

\section{Connection between numerical and physical values}

\label{sec:numerical_physical}

The field equations solved in Sec.~\ref{Sec:solutions} are written in terms of the dimensionless parameters $\kappa$, $g$, $g'$, and $\beta$ . The  numerical solutions are therefore scale-free:
the same dimensionless configuration can describe physical objects of vastly
different masses and sizes, depending on the units used to restore
dimensions. In this section, we make the conversion explicit.

In our setup, the conversion involves two logically independent inputs. First, once
$\bm G$ and $\bm c$ are fixed, the dimensionless gravitational coupling $\kappa$ determines
the Higgs vev $\bm\Phi_0$. Second, the chosen physical
W-boson rest energy ${\bm m}_{\rm W}{\bm c^2}$ determines the dimensionful gauge coupling $\bm g$ and,
with it, the length and mass units of the solutions. Throughout this section,
$\bm G$, $\bm c$, and $\bm\hbar$ are kept at their measured CODATA
values~\cite{Mohr:2024kco}, while the dimensionless electroweak ratios are fixed to their Standard Model values used in the numerical solutions.

We follow the notation introduced above: boldface symbols
denote dimensionful quantities, whereas ordinary symbols denote dimensionless quantities.
In particular,
\begin{equation}
    g=\frac{\bm g}{\bm g_0},\qquad
    g'=\frac{\bm g'}{\bm g_0}\qquad\text{with}\qquad\bm g_0=\sqrt{\bm g^2+\bm g'^2}.
\end{equation}

The parameters \(g\) and \(g'\) are therefore ratios, while \(\bm g\) and \(\bm g'\)
are the dimensionful gauge couplings. Using the CODATA values quoted in Ref.~\cite{Mohr:2024kco}, we take,
\begin{equation}
    {\bm c}=299\,792\,458\,{\rm m\,s^{-1}},
    \qquad
    {\bm\hbar}=6.582\,119\,569\times10^{-16}\,{\rm eV\,s},
    \label{eq:num_ch}
\end{equation}
and, 
\begin{equation}
    \bm G
    =
    6.674\,30\times 10^{-11}\,
    {\rm m^3\,kg^{-1}\,s^{-2}}
    =
    1.069\,34\times 10^{-29}\,
    {\rm m^5\,eV^{-1}\,s^{-4}},
    \label{eq:num_G}
\end{equation}
where we used $1\,\mathrm{eV}=1.602\,176\,634\times10^{-19}\,\mathrm{J}$.

The dimensionless gravitational coupling used in our numerical calculations is,
\begin{equation}
    \kappa=
    \frac{8\pi \bm G \bm{\Phi}_0^2}{\bm c^4}.
    \label{eq:kappa_sec5}
\end{equation}

Hence, once \(\kappa\) is fixed, the dimensionful Higgs vev is,
\begin{equation}\label{eq:Phi0}
    \bm{\Phi}_0
    =
    \sqrt{\frac{\kappa {\bm c}^4}{8\pi \bm G}}.
\end{equation}

Taking $\kappa=2$, the reference value used for the solutions S1--S3, gives,
\begin{equation}
    \bm{\Phi}_0
    =
    7.753\times10^{30}\,
    \left({\rm eV}/{\rm m}\right)^{1/2}=
    3.103\times10^{21}\,
    \left({\rm J}/{\rm m}\right)^{1/2}.
    \label{eq:num_phi}
\end{equation}

Passing to the convention of Ref.~\cite{PhysRevD.110.030001} yields the value $v\equiv\sqrt{2\bm{\hbar}\bm{c}}\times \bm{\Phi}_0=4.871\times 10^{18}\,{\rm GeV}$, which is clearly far above the Standard Model value ($v=246.2\,{\rm GeV}$). As previously discussed, the present model should be interpreted as a dark electroweak sector that preserves the gauge-symmetry structure of the Weinberg-Salam theory while allowing its characteristic energy scale to differ radically from that of the observed electroweak sector. Using the Standard Model value of $\bm \Phi_0$ in Eq.~\eqref{eq:kappa_sec5} would lead to $\kappa\sim 10^{-33}$, as discussed in Sec.~\ref{sec:model}. Taking $\kappa$ of order unity instead places the dark Higgs vev just below the Planck scale and therefore, the gravitational interaction and electroweak self-interactions are of the same order of magnitude.

The length and energy scales associated with the dimensionless variables are,
\begin{equation}
    {\bm\ell}_0=\frac{1}{\bm g_0\bm{\Phi}_0},
    \qquad
    {\bm m}_0{\bm c}^2=\frac{\bm\hbar \bm c}{\bm\ell_0}
    =
    {\bm\hbar \bm c}\,\bm g_0\bm{\Phi}_0.
    \label{eq:natural_scale}
\end{equation}

Therefore, a dimensionless boson mass \(m_{\rm num}\) corresponds to the physical rest energy,
\begin{equation}
    {\bm m}{\bm c}^2=m_{\rm num}\,\bm m_0 \bm c^2 .
    \label{eq:num_to_phys}
\end{equation}

In the vacuum, the dimensionless boson masses are,
\begin{equation}
    m_{\rm Z}=\frac{1}{\sqrt2},\qquad
    m_{\rm W}=\frac{g}{\sqrt2},\qquad
    m_{\rm H}=\sqrt{\frac{\beta}{2}}.
        \label{eq:boson_mass_bis}
\end{equation}

Therefore, the physical $W$-boson rest energy is,
\begin{equation}\label{eq:mw}
    \bm m_{\rm W}\bm c^2
    =
    \frac{g}{\sqrt2}\,\bm m_0 \bm c^2
    =
    \frac{\bm\hbar \bm c\,\bm g\,\bm{\Phi}_0}{\sqrt2},
\end{equation}
where in the last equality, we used \(\bm g=g\bm g_0\). Hence,
\begin{equation}\label{eq:g}
    \bm g
    =
    \frac{\sqrt2\,\bm m_{\rm W}\bm c^2}
         {\bm\hbar \bm c\,\bm{\Phi}_0}.
\end{equation}

Now, we keep the dimensionless electroweak ratios at their Standard Model values,
\begin{equation}
    g'=\sin\theta_{\rm W}\simeq\sqrt{0.223},
    \qquad
    g=\cos\theta_{\rm W}\simeq\sqrt{1-0.223},\qquad\beta=1.88,
\end{equation}
and set the (dark) $W$-boson mass to the value inspired by the Proca star interpretation of GW190521, in which the signal is reproduced by the head-on merger of two Proca stars composed of an ultralight vector boson with mass $\simeq8.7\times10^{-13}\,\mathrm{eV}/\bm{c}^2$, collapsing into a black hole~\cite{CalderonBustillo:2020fyi,CalderonBustillo:2022cja}. We therefore
take,
\begin{equation}
   \bm m_{\rm W}\bm c^2=8.7\times10^{-13}\,{\rm eV},
\end{equation}
which yields,
\begin{equation}
    \bm g
    =
    8.042\times10^{-37}\,
    ({\rm eV\,m})^{-1/2}.
\end{equation}

This mass should be regarded only as a convenient and astrophyiscally-motivated choice for fixing the overall scale of the solutions. The Proca field used in Refs.~\cite{CalderonBustillo:2020fyi,CalderonBustillo:2022cja} 
is neutral and free, whereas the $W$ field considered here is electrically charged and interacts with the electromagnetic, Z, and Higgs fields. The two systems are therefore dynamically distincts, and no direct correspondence between their merger dynamics can be inferred from the common boson mass. Establishing whether the mergers of the present bosonic stars could produce a similar signal would require a dedicated dynamical study.

According to Eqs.~\eqref{eq:num_to_phys} and \eqref{eq:boson_mass_bis}, the dimensionless parameter $g$ can also be interpreted as the ratio between the masses of the $W$ and $Z$ bosons,
\begin{equation}
    g=\frac{\bm m_\text{W}}{\bm m_\text{Z}},
\end{equation}
and likewise, one has,
\begin{equation}
    \beta=\left(\frac{\bm m_\text{H}}{\bm m_\text{Z}}\right)^2=g^2\left(\frac{\bm m_\text{H}}{\bm m_\text{W}}\right)^2.
\end{equation}

From these relations, the remaining dark-sector boson masses are,
\begin{equation}\label{eq:masses}
    \bm m_{\rm Z}\bm c^2=9.870\times10^{-13}\,{\rm eV},\qquad
    \bm m_{\rm H}\bm c^2=1.353\times10^{-12}\,{\rm eV}.
\end{equation}

The remaining dimensionful gauge couplings follow from,
\begin{equation}
    \bm g_0=\frac{\bm g}{g},
    \qquad
    \bm g'=g'\bm g_0.
\end{equation}

Numerically,
\begin{equation}
    \bm g_0
    =
    9.123\times10^{-37}\,
    ({\rm eV\,m})^{-1/2}
    \qquad\text{and}\qquad
    \bm g'
    =
    4.308\times10^{-37}\,
    ({\rm eV\,m})^{-1/2}.
    \label{eq:num_g0}
\end{equation}

Thus, for \(\kappa=2\) and \(\bm m_{\rm W}\bm c^2=8.7\times10^{-13}\,{\rm eV}\), the physical
parameters entering the dimensionful action \eqref{eq:action-dimensional} are, 
\begin{equation}
    \bm{\Phi}_0
    =
     7.753\times10^{30}\,
    ({\rm eV}/{\rm m})^{1/2},
\end{equation}
and
\begin{equation}
    \bm g
    =
    8.042\times10^{-37}\,
    ({\rm eV\,m})^{-1/2},
    \qquad
    \bm g'
    =
    4.308\times10^{-37}\,
    ({\rm eV\,m})^{-1/2}.
\end{equation}

 Having specified the dimensionful model parameters above, we now explain  how to convert the dimensionless ADM mass of our numerical solutions into its physical value. The metric coefficient $g_{tt}$ is dimensionless and may be expressed equivalently in terms of dimensionful or dimensionless quantities; its asymptotic expansion reads,
\begin{equation}
    -g_{tt}=1-\frac{2M}{r}+\dots=1-\frac{2\bm{GM}}{\bm{c}^2\bm{r}}+\dots,
\end{equation}
where $\bm{M}$ and $\bm{r}$ are the dimensionful mass and radial coordinate, respectively. Identifying the two expressions and using $\bm{r}=r\,\bm{\ell}_0$, one obtains the relation,
\begin{equation}\label{eq:Madm}
    \bm{M}=\frac{\bm{c}^2}{\bm{G}}\frac{\bm{r}}{r}M=\frac{\bm{c}^2\bm{\ell}_0}{\bm{G}}M \equiv \bm{M}_0 M.
\end{equation}

Hence, the mass and length units are independent of $\kappa$ and depend only on the chosen physical $W$-boson mass, once $g$ is fixed:
\begin{equation}
    \bm{\ell}_0
    = \frac{1}{{\bm g}_0{\bm\Phi}_0}=
    \frac{g}{\sqrt{2}}\frac{\bm\hbar\bm c}{\bm m_{\rm W}\bm c^2},
    \qquad
    \bm{M}_0
    =
    \frac{g}{\sqrt{2}}
    \frac{\bm\hbar\bm c^3}{\bm G\,\bm m_{\rm W}\bm c^2}.
    \label{eq:scales_mw}
\end{equation}

This does not mean that the physical solutions are independent of $\kappa$: their dimensionless masses $M(\kappa)$ and effective radii $R_{99}(\kappa)$ do depend on the gravitational coupling, as seen in Sec.~\ref{Sec:solutions}. Equation~\eqref{eq:scales_mw} only states that the factors converting a given dimensionless mass or radius into physical units are independent of $\kappa$.

For the reference value $\bm m_{\rm W}\bm c^2=8.7\times10^{-13}\,\mathrm{eV}$, the corresponding ADM mass unit is,\begin{equation}\label{eq:M}
    {\bm M}_0=\frac{\bm{c}^2\bm{\ell}_0}{\bm{G}}=1.904\times 10^{32}\,\text{kg}\simeq95.76\,\bm{M}_\odot,
\end{equation}
where $\bm{M}_\odot=1.988\times 10^{30}\,\text{kg}$ is the solar mass. We also give the value of the length unit,
\begin{equation}\label{eq:length}
    \bm\ell_0
    =\frac{1}{\bm g_0\bm{\Phi}_0}
    =1.414\times10^2\,\mathrm{km},
\end{equation}
which is useful for converting the dimensionless $R_{99}$ into its physical value via $\bm{R}_{99}=R_{99}\,\bm{\ell}_0$.

As a concrete application, consider the three 
reference solutions
considered in Sec.~\ref{Sec:solutions}. Their physical masses $\bm M$
and effective radii ${\bm R}_{99}$ are,
\begin{align}
    {\rm S1}:\quad&
    \bm M\simeq245\,\bm M_\odot,
    \qquad\qquad    
    \bm R_{99}\simeq3.32\times10^3\,{\rm km},
    \notag
    \\
    {\rm S2}:\quad&
    \bm M\simeq131\,\bm M_\odot,
    \qquad\qquad    
    \bm R_{99}\simeq1.52\times10^3\,{\rm km},
    \notag
    \\
    {\rm S3}:\quad&
    \bm M\simeq123\,\bm M_\odot,
    \qquad\qquad    
    \bm R_{99}\simeq2.10\times10^3\,{\rm km}.
    \label{eq:reference_physical_values}
\end{align}

As another example, consider the family of solutions where $\kappa$ is varied shown in Fig.~\ref{fig_varkap}. At the critical gravitational coupling value $\kappa_c$, the maximum mass in physical units is ${\bm M}\simeq 7416\,\bm M_\odot$. At the same time, the effective radius drops from $\sim 70\,000\,{\rm km}$ to $\sim 4\,500\,{\rm km}$.
\begin{table}[h]
\centering
\small
\renewcommand{\arraystretch}{1.25}
\begin{tabularx}{\textwidth}{@{}cY@{\hspace{0pt}\vrule\hspace{0pt}}YYYY@{}}
\toprule
\shortstack{$\kappa$\\\phantom{t}}
& \shortstack{$\bm m_{\rm W}\bm c^2$\\(eV)}
& \shortstack{$\bm m_{\rm H}\bm c^2$\\(eV)}
& \shortstack{$\bm{\Phi}_0$\\$\left(\sqrt{\mathrm{eV}/\mathrm{m}}\right)$}
& \shortstack{$\bm{\ell}_0$\\(km)}
& \shortstack{$\bm M_0$\\$(M_\odot)$} \\
\midrule
1 & $1.0\times10^{-10}$ & $1.6\times10^{-10}$
  & $5.5\times10^{30}$ & $1.23$ & $0.83$ \\
1 & $8.7\times10^{-13}$ & $1.4\times10^{-12}$
  & $5.5\times10^{30}$ & $1.41\times10^{2}$ & $95.7$ \\
1 & $1.0\times10^{-20}$ & $1.6\times10^{-20}$
  & $5.5\times10^{30}$ & $1.23\times10^{10}$ & $8.3\times10^{9}$ \\
2 & $1.0\times10^{-10}$ & $1.6\times10^{-10}$
  & $7.8\times10^{30}$ & $1.23$ & $0.83$ \\
2 & $8.7\times10^{-13}$ & $1.4\times10^{-12}$
  & $7.8\times10^{30}$ & $1.41\times10^{2}$ & $95.7$ \\
2 & $1.0\times10^{-20}$ & $1.6\times10^{-20}$
  & $7.8\times10^{30}$ & $1.23\times10^{10}$ & $8.3\times10^{9}$ \\
\bottomrule
\end{tabularx}
\caption{Summary of physical scales. The solutions are computed within a
dimensionless scheme and therefore exist independently of the overall
physical scale, which we fix through the value of the $W$-boson mass.
As emphasized in Sec.~\ref{Sec:solutions}, ultralight $W$ bosons yield
condensates with astrophysical masses and sizes, whereas heavier ones
would correspond to microscopic objects. We illustrate this feature by
comparing, for two illustrative values of $\kappa$ and three values of
$\bm m_{\rm W}\bm c^2$, the corresponding physical values of the Higgs
mass $\bm m_{\rm H}\bm c^2$, the Higgs vev
$\bm{\Phi}_0$, the length scale $\bm{\ell}_0$ converting dimensionless
lengths into physical units, and the mass scale
${\bm M}_0\equiv\bm c^2\bm{\ell}_0/\bm G$ converting the dimensionless
mass into physical units given in solar masses. Note that
$\bm{\Phi}_0$ is determined by $\kappa$ alone, whereas
$\bm{\ell}_0$ and ${\bm M}_0$ depend only on
$\bm m_{\rm W}$.}
\label{tab:physical_scales}
\end{table}

Table~\ref{tab:physical_scales} summarizes the physical scales discussed in this 
section. 
The numerical solutions presented in Sec.~\ref{Sec:solutions}
exist for any nonzero value of the $W$-boson mass $\bm{m}_{\mathrm{W}}$, and the conversion 
to physical units proceeds in two independent steps: the gravitational 
coupling $\kappa$ determines the Higgs vev
$\bm{\Phi}_0$ through Eq.~\eqref{eq:Phi0}, while $\bm{m}_{\mathrm{W}}$ 
determines the dimensionful gauge coupling $\bm{g}$ through 
Eq.~\eqref{eq:g} and, with it, the length and mass units in Eq.~\eqref{eq:scales_mw}. Both units 
are inversely proportional to the boson mass and independent of $\kappa$ (although the solutions themselves do depend on $\kappa$). The table illustrates this 
for two values of $\kappa$ and three values of 
$\bm{m}_{\mathrm{W}}\bm{c}^2$, listing the Higgs mass, the Higgs vev, and the mass and length units.

The inverse dependence of $\bm{M}_0$ (and ${\bm\ell}_0$) on $\bm m_{\rm W}$ is the familiar boson star scaling. For configurations with $M=\mathcal{O}(1)$, the interval $\bm m_{\rm W}\bm c^2\sim10^{-20}$--$10^{-10}\,\mathrm{eV}$ corresponds to mass scales ranging from approximately one solar mass to $10^{10}\,\bm{M}_\odot$. The specific value 
$\bm{m}_{\mathrm{W}}\bm{c}^2=8.7\times10^{-13}\,$eV, 
coming from the Proca star interpretation of GW190521~\cite{CalderonBustillo:2022cja,CalderonBustillo:2020fyi}, yields to the value $\sim 10^2\,\bm{M}_\odot$. 

\section{Further Remarks}\label{sec:remarks}

The spherically symmetric solutions constructed in this work demonstrate that the Einstein-Weinberg-Salam theory admits everywhere regular, asymptotically flat, electrically charged self-gravitating condensates. The massive $W$, $Z$, and Higgs fields are localized within the condensate, whereas the massless ${\rm U}(1)_{\rm em}$ field produces the Coulomb tail carrying the total electric charge originating from the charged $W$ bosons. Fixing the value of the dimensionless gravitational coupling $\kappa$, the solutions form a one-parameter family parametrized by the $W$-field frequency $\omega$ and retain the qualitative structure familiar from bosonic stars: they emerge continuously from the vacuum as $\omega\to m_{\rm W}$, reach a maximal mass, and subsequently exhibit the characteristic spiral behavior in parameter space. 

The solution families exist for arbitrarily large gravitational coupling $\kappa$, but shrink as $\kappa$ is lowered and cease to exist slightly below the critical coupling $\kappa_c=4\sin^2\theta_{\rm W}$. In particular, for the Standard Model value of the weak mixing angle $\theta_{\rm W}$ considered in this work, the flat-space limit could not be reached. Our results demonstrate that the critical coupling $\kappa_c$ plays an important role, marking a transition between two qualitatively different regimes. Heuristically, it corresponds to the gravitational coupling value at which the long-range gravitational attraction and electric repulsion balance. Above $\kappa_c$, solutions with $\omega\approx m_{\rm W}$ are highly dilute, close to the trivial vacuum, and the electroweak interactions within the configuration essentially reduce to ordinary electromagnetism, with the Higgs field remaining close to its vev. Below $\kappa_c$, the configurations become more compact and increasingly nonlinear, with the Higgs deviating significantly from its vev. Notably, the mass-to-charge ratio approaches the extremal black hole value closely at $\kappa_c$. These features are reported here for the first time, though similar mechanisms may arise in other bosonic star models. 

Bosonic stars have been studied for the past six decades, primarily because of their viability as dark matter candidates and their capacity to act as black hole mimickers. If dark matter contains ultralight bosonic degrees of freedom, these fields may form macroscopic self-gravitating configurations whose mass and size are controlled primarily by the boson mass, but also by the structure of the underlying field theory.  Their existence, stability, merger dynamics, and possible gravitational-wave signatures could encode information about physics beyond the Standard Model, making strong-gravity systems indirect particle physics laboratories. Although direct observational evidence of bosonic stars remains elusive, their study is therefore justified even if such objects are not ultimately realized in Nature. Nevertheless, to elevate these field configurations from theoretical curiosities to physical realities, three stringent criteria must be met: they must emerge naturally from a well-motivated fundamental theory, possess a viable astrophysical formation mechanism, and exhibit sufficient dynamical stability to survive environmental perturbations.

The present work makes progress mainly toward the first of these requirements. In contrast with the free Proca model, the vector boson mass arises via a Higgs mechanism, and the $W$ condensate is accompanied self-consistently by the Higgs, $Z$, and a massless gauge field -- precisely the bosonic content of the electroweak model. However, astrophysically relevant configurations cannot be realized using the observed electroweak energy scale. Our model should therefore be regarded as a dark electroweak sector with a Standard-Model-like gauge structure and mass ratios, rather than as the physical Standard Model shifted as a whole to a lower energy scale~\cite{Freitas:2021cfi}. In this precise sense, the construction inherits a matter structure known to occur in Nature, while its ultralight realization remains hypothetical.

For the dimensionless model parameters explored here, and adopting the ultralight dark $W$-boson mass $8.7\times10^{-13}\,{\rm eV}/c^2$, motivated by the analyses in Refs.~\cite{CalderonBustillo:2022cja,CalderonBustillo:2020fyi}, the bosonic stars can reach masses of order $100\,M_\odot$, and up to $\sim 1000\,M_\odot$ near the critical value of the gravitational coupling. Their compactness can approach -- but never reach -- the Schwarzschild black hole value. From these first principles, the stars are thus likely to be able to mimic black holes in the intermediate-mass range. At the same time, since the mass scale of the solutions is inversely proportional to the $W$-boson mass, choosing values across the ultralight range $10^{-20}-10^{-10}\,\rm{eV}/c^2$ yields objects with a variety of masses, ranging from stellar to supermassive black hole values.

\medskip

This work opens new perspectives for the study of bosonic stars by embedding them in a gauge structure that we know to exist in Nature, providing a self-consistent field-theoretic framework for exploring compact objects at the interface of high-energy physics and strong gravity. Looking forward, several natural extensions of this work present themselves.

First, in the static case, pure Proca stars are known to possess a non-spherical ground state, with spherical stars being unstable and decaying into prolate field configurations~\cite{Herdeiro:2023wqf,Herdeiro:2026agu,Diez-Tejedor:2026fnc}. Although the dark $W$ boson considered in this work shares several similarities with the free Proca field, the additional degrees of freedom present in the theory may or may not help stabilize the spherically symmetric configuration. Nonetheless, the vector nature of the $W$ field suggests that the ground state might also depart from spherical symmetry. Consequently, constructing static solutions beyond spherical symmetry and determining which configuration corresponds to the true ground state of the theory remain open problems, which we plan to investigate in future work. Second, from an astrophysical perspective, compact objects are generically expected to possess angular momentum. Generalizing the present static solutions to include rotation is therefore another important step. Third, we plan to investigate whether the present solitons admit static, spherically symmetric black hole generalizations. The existence of black holes with non-Abelian hair is well-established~\cite{VolkovGaltsov1989,KuenzleMasood1990,Bizon1990,KleihausKunz1997,KleihausKunz2001,BreitenlohnerForgacsMaison1992,Greene:1992fw}. More directly related to this work are the magnetically charged black holes with electroweak hair constructed within the Einstein--Weinberg--Salam theory~\cite{Bai2021,Gervalle2024,Gervalle2025}. These examples demonstrate that nontrivial gauge and Higgs configurations can coexist with a regular event horizon. Their nontrivial gauge-field structure, however, involves the magnetic sector, whereas the solitons studied here are purely electric. Whether the coupled $W$, $Z$, Higgs and electromagnetic fields can support a localized, purely electric condensate outside a static horizon therefore remains to be investigated.

Finally, although our focus has been on self-gravitating configurations within the Einstein-Weinberg-Salam theory, we have also numerically constructed flat-spacetime solitons with $\kappa=0$ for suitable choices of the electroweak parameters. They are not presented here, however, since we were unable to construct them while keeping the electroweak mass ratios fixed at their physical values. Whether such solitons are genuinely absent for the Standard Model parameters or instead exist on another disconnected branch of solutions therefore remains to be determined.\footnote{As this work was being completed, we became aware of the independent construction and study of flat-spacetime solitons in Ref.~\cite{EWballs}. The authors consider both electric-type and magnetic-type \emph{electroweak balls}, the former being the flat-spacetime analogues of the spherically symmetric electric configurations constructed here. Within the families explored, they likewise find no solutions at the measured Standard Model values of the electroweak mass ratios. While this does not establish non-existence, this independent
observation further motivates determining whether such non-topological solitons exist in the Standard Model and, if not, identifying the physical mechanism that prevents their existence.}

\subsection*{Acknowledgements}
The authors are grateful to Vinícius Oliveira for his insightful discussions. E.S.C.F. is supported by CNPq under the Postdoctoral Junior (PDJ) program, grant 153723/2025-4. This work was partially supported by FCT I.P. under Project 2025.09655.CPCA.A2 através da FCCN at Deucalion supercomputer, jointly funded by EuroHPC JU and Portugal.

\appendix


\section{A mass formula}
\label{sec:massformula}

We close the analysis 
by deriving a Smarr-type
mass formula \cite{herdeiro2023procahiggs} for the static, horizonless configurations considered here.
We have an everywhere timelike Killing vector $\xi$, associated in
adapted coordinates with the time coordinate $t$. Since the spacetime is static and the neutral fields $Z_\mu$ and $\phi$
are real, they inherit the symmetry generated by the timelike Killing
vector. Therefore,
\begin{equation}
    \mathcal{L}_{\xi}Z_{\mu}=0\,,\qquad \mathcal{L}_{\xi}\phi=0\,.
    \label{eq:lie-neutral}
\end{equation}

The charged sector, however, needs to be invariant only up to a residual
${\rm U}(1)$ gauge transformation. Therefore, without fixing the
gauge, the most general conditions are,
\begin{equation}
    \mathcal{L}_{\xi}A_{\mu}=\frac{1}{e}\nabla_{\mu}\chi,
    \qquad
    \mathcal{L}_{\xi}w_{\mu}=-i\chi w_{\mu},
    \label{eq:lie-charged}
\end{equation}
where $\chi$ is a general, spacetime-dependent function. Under the gauge transformation~\eqref{eq:residualU1}, this function transforms as,
\begin{equation}
    \chi\longrightarrow\chi+\mathcal{L}_{\xi}\lambda.
\end{equation}

It is then useful to introduce the gauge-invariant quantity,
\begin{equation}
    X\equiv\frac{\chi}{e}-\xi^{\nu}A_{\nu}.
    \label{eq:def-X}
\end{equation}

The above symmetry conditions then imply,
\begin{align}
    \nabla_{\mu}\left(\xi^{\nu}Z_{\nu}\right)
    &=
    \xi^{\nu}Z_{\mu\nu},
    \label{eq:static-Z-identity}
    \\
    \nabla_{\mu}X
    &=
    \xi^{\nu}F_{\nu\mu},
    \label{eq:static-A-identity}
    \\
    {\cal D}_{\mu}\left(\xi^{\nu}w_{\nu}\right)
    &=
    \xi^{\nu}w_{\mu\nu}
    -i\left[
        eX+g^{2}\left(\xi^{\nu}Z_{\nu}\right)
    \right]w_{\mu},
    \label{eq:static-w-identity}
\end{align}
together with the complex conjugate of
Eq.~\eqref{eq:static-w-identity}. The first two equations follow from
Cartan's identity, while the last one follows from the definition of
$w_{\mu\nu}$ in Eq.~\eqref{eq:w-proca-field}. Contracting
Eq.~\eqref{eq:static-A-identity} with $\xi^\mu$ also gives
$\mathcal{L}_{\xi}X=0$. The goal is then to compute the Komar formula~\eqref{eq:mass_integral},
\begin{equation}
    \mathcal{M}
    =
    -2\int_{\Sigma}
    \left(
        T^{\mu}{}_{\nu}
        -\frac{1}{2}\delta^{\mu}_{\nu}T
    \right)
    \xi^{\nu}\,d\Sigma_{\mu},
    \label{eq:komar-mass}
\end{equation}
in terms of physical quantities. From the Lagrangian~\eqref{eq:Lagr}, the energy-momentum tensor can be
written as,
\begin{equation}
    T_{\mu\nu}
    =
    -\frac{2}{\sqrt{-{\rm g}}}
    \frac{\delta\!\left(\sqrt{-{\rm g}}\,\mathcal{L}_{\rm WS}\right)}
         {\delta {\rm g}^{\mu\nu}}
    =
    -2\frac{\partial\mathcal{L}_{\rm WS}}
             {\partial {\rm g}^{\mu\nu}}
    +{\rm g}_{\mu\nu}\mathcal{L}_{\rm WS}.
    \label{eq:Tfromvar}
\end{equation}
Using the field equations~\eqref{eq:A}--\eqref{eq:phi}, together with
Eqs.~\eqref{eq:static-Z-identity}--\eqref{eq:static-w-identity}, one
obtains the on-shell identity,
\begin{equation}
    -2\frac{\partial\mathcal{L}_{\rm WS}}
             {\partial {\rm g}^{\mu\nu}}\xi^{\nu}
    =
    \nabla^{\sigma}\mathcal{V}_{\mu\sigma},
    \label{eq:gauge-independent-reduction}
\end{equation}
where,
\begin{align}
    \mathcal{V}_{\mu\sigma}
    ={}&
    \left(\xi^{\nu}Z_{\nu}\right)
    \left(-Z_{\mu\sigma}+g^{2}\psi_{\mu\sigma}\right)-\frac{1}{2}
    \left[
       \bar w_{\mu\sigma}\left(\xi^{\nu}w_{\nu}\right)
       +w_{\mu\sigma}\left(\xi^{\nu}\bar w_{\nu}\right)
    \right]
    +\left(F_{\mu\sigma}+e\psi_{\mu\sigma}\right)X.
    \label{eq:gauge-independent-potential}
\end{align}
All terms in $\mathcal{V}_{\mu\sigma}$ are gauge invariant, and
$\mathcal{V}_{\mu\sigma}$ is antisymmetric. It follows from Eqs.~\eqref{eq:Tfromvar} and
\eqref{eq:gauge-independent-reduction} that,
\begin{equation}
    T^{\mu}{}_{\nu}\xi^{\nu}
    =
    \nabla_{\sigma}\mathcal{V}^{\mu\sigma}
    +\mathcal{L}_{\rm WS}\xi^{\mu}.
    \label{eq:stress-reduction}
\end{equation}
Moreover, the trace of Einstein's equations~\eqref{eq:Ein} gives,
\begin{equation}
    R=-\kappa T.
    \label{eq:einstein-trace-mass}
\end{equation}
Consequently,
\begin{equation}
    \left(
        T^{\mu}{}_{\nu}
        -\frac{1}{2}\delta^{\mu}_{\nu}T
    \right)\xi^\nu
    =
    \nabla_{\sigma}\mathcal{V}^{\mu\sigma}
    +
    \left(
        \frac{1}{2\kappa}R+\mathcal{L}_{\rm WS}
    \right)\xi^\mu.
    \label{eq:komar-integrand-reduction}
\end{equation}
Substitution into Eq.~\eqref{eq:komar-mass} yields,
\begin{equation}
    \mathcal{M}
    =
    -2\int_{\Sigma}
    \nabla_{\sigma}\mathcal{V}^{\mu\sigma}\,d\Sigma_{\mu}
    -2L,
    \label{eq:mass-before-boundary}
\end{equation}
where the on-shell spatial Lagrangian is,
\begin{equation}
    L
    \equiv
    \int_{\Sigma}
    \left(
        \frac{1}{2\kappa}R+\mathcal{L}_{\rm WS}
    \right)
    \xi^{\mu}\,d\Sigma_{\mu}.
    \label{eq:def-on-shell-L}
\end{equation}

Since the solutions are 
horizonless, there is no inner boundary. At spatial infinity, the massive fields $Z_{\mu}$ and $w_{\mu}$ vanish exponentially fast, and 
so does $\psi_{\mu\nu}$. Moreover, the electromagnetic field admits the Coulombian asymptotic behavior $F_{tr}=\mathcal{O}(r^{-2})$, and it follows directly that $X=X_{\infty}+\mathcal{O}(r^{-1})$. Using this in Eq.~\eqref{eq:mass-before-boundary} yields the Smarr-type mass formula,
\begin{equation}
        \mathcal{M}=8\pi X_{\infty}Q_e- 2 L\,.
\end{equation}

After deriving the gauge-invariant form of the mass formula, it is useful to bring it to the gauge we are using in the present work where the potential $A_\mu$ vanishes at infinity and $\chi=\omega$. Therefore, we have,
\begin{equation}
        \mathcal{M}=\dfrac{8\pi\omega}{e}Q_e- 2 L\,,
\end{equation}
and the corresponding differential mass formula is,
\begin{equation}\label{Eq:diffmass}
    \delta  \mathcal{M}=\dfrac{4\pi\omega}{e}\delta Q_e\,.
\end{equation}

The Smarr formula also provides an independent test of the numerical solutions by allowing a direct comparison between $\mathcal M$ and $8\pi X_\infty Q_e-2L$. Along each one-parameter family constructed at fixed couplings, the differential relation~\eqref{Eq:diffmass} can additionally be tested using finite differences between neighboring solutions.

\bibliographystyle{hhieeetr}
\bibliography{biblio}

\end{document}